\documentclass[11pt,a4paper]{article}
\pdfoutput=1
\usepackage{amssymb,amsmath,amsfonts, mathtools, mathrsfs}
\usepackage[utf8]{inputenc} % per usare le lettere accentate come à,è etc...
\usepackage[dvipsnames]{xcolor}

\newif\ifnatbibsort\natbibsorttrue

\relax

\ifnatbibsort\RequirePackage[numbers,sort&compress]{natbib}\else\RequirePackage[numbers,compress]{natbib}\fi
\RequirePackage[colorlinks=true
,urlcolor=blue
,anchorcolor=blue
,citecolor=blue
,filecolor=blue
,linkcolor=blue
,menucolor=blue
,pagecolor=blue
,linktocpage=true
,pdfproducer=medialab
,pdfa=true
]{hyperref}

\usepackage{graphicx}
\usepackage{caption}
\usepackage{tensor}
\usepackage{subfigure}
\usepackage{enumerate}
\usepackage{dsfont}
\usepackage{verbatim}
\usepackage{amsfonts,euscript,amssymb,stmaryrd,braket}
\usepackage{tikz}
\usetikzlibrary{arrows,decorations.markings,patterns}
\usepackage{slashed}

\def\clock{{\count0=\time
		\divide\count0 60
		\ifnum\count0<10 0\fi\the\count0
		\multiply\count0 -60 \advance\count0 \time
		:\ifnum\count0<10 0\fi \the\count0
}}
\newcommand{\timestamp}{{\small\vbox{\hbox{\tt\jobname.tex}
			\hbox{\the\day/\the\month/\the\year, \clock}}}}

\newcommand{\bea}{\begin{eqnarray}}
\newcommand{\eea}{\end{eqnarray}}

\newcommand{\be}{\begin{equation}}
\newcommand{\ee}{\end{equation}}

\makeatletter
\let\old@startsection=\@startsection
\let\oldl@section=\l@section
\renewcommand{\@startsection}[6]{\old@startsection{#1}{#2}{#3}{#4}{#5}{#6\mathversion{bold}}}
\renewcommand{\l@section}[2]{\oldl@section{\mathversion{bold}#1}{#2}}
\makeatother

\numberwithin{equation}{section}
\def \tpsi {{\widetilde \psi}}

\def \tx {\widehat x} 
\def \tPsi {\widetilde \Psi}

\usepackage{color}

\def \RR {{\mathbb R}}

\def \SS {{\mathbb S}}
\def \U {{\cal U}}

\def\ri {{\rm i}}
\def\rd {{\rm d}}
\def\e {{\rm e}}
\def\prt {{\partial}}

\begin{document}
	\renewcommand{\thefootnote}{\arabic{footnote}}

	\overfullrule=0pt
	\parskip=2pt
	\parindent=12pt
	\headheight=0in \headsep=0in \topmargin=0in \oddsidemargin=0in

	\vspace{ -3cm} \thispagestyle{empty} \vspace{-1cm}
	\begin{flushright} 
		\footnotesize
		\textcolor{red}{\phantom{print-report}}
	\end{flushright}

\begin{center}
	\vspace{.5cm}

	{\Large\bf \mathversion{bold}
	Modular Hamiltonians for the massless Dirac field
	}
	\\
	\vspace{.25cm}
	\noindent
	{\Large\bf \mathversion{bold}
	in the presence of a defect}

	\vspace{0.8cm} {
		Mihail Mintchev$^{\,a,}$\footnote[1]{mintchev@df.unipi.it}
		and Erik Tonni$^{\,b,}$\footnote[2]{erik.tonni@sissa.it}
	}
	\vskip  0.7cm
	
	\small
	{\em
		$^{a}\,$Dipartimento di Fisica, Universit\'a di Pisa and INFN Sezione di Pisa, \\
		largo Bruno Pontecorvo 3, 56127 Pisa, Italy
		\vskip 0.05cm
		$^{b}\,$SISSA and INFN Sezione di Trieste, via Bonomea 265, 34136, Trieste, Italy 
	}
	\normalsize

\end{center}

\vspace{0.3cm}
\begin{abstract} 
\noindent
We study the massless Dirac field 
on the line in the presence of a point-like defect characterised by a unitary scattering matrix, 
that allows both reflection and transmission.
Considering this system in its ground state, 
we derive the modular Hamiltonians of the subregion given by the union of two disjoint equal intervals at the same distance from the defect.
The absence of energy dissipation at the defect implies the existence of two phases, where either the vector or the axial symmetry is preserved.
Besides a local term, the densities of the modular Hamiltonians contain also a sum of scattering dependent bi-local terms, which 
involve two conjugate points generated by the reflection and the transmission. 
The modular flows of each component of the Dirac field 
mix the trajectory passing through a given initial point 
with the ones passing through its reflected and transmitted conjugate points. 
We derive the two-point correlation functions along the modular flows
in both phases and show that they satisfy the Kubo-Martin-Schwinger condition.
The entanglement entropies are also computed, finding that they do not depend on the scattering matrix. 
\end{abstract}

\newpage

%%%%%%%%%%%%%%%%%%%%%%%%%%%%%%%%%%%%%
\tableofcontents
%%%%%%%%%%%%%%%%%%%%%%%%%%%%%%%%%%%%%

%\newpage
%%%%%%%%%%%%%%%%%%%%%%%%%%%%%%%%%%%%%%%%%%%
\section{Introduction}
\label{sec_intro}

The study of the geometric entanglement between complementary spatial regions 
has provided important insights in 
quantum field theory, quantum gravity, condensed matter and quantum information 
during the last few decades.

Considering a quantum system in the state described by a density matrix $\rho$
and assuming that its Hilbert space is factorised as $\mathcal{H} = \mathcal{H}_A \otimes \mathcal{H}_B$
in correspondence with the spatial bipartition $A \cup B$,
the reduced density matrix $\rho_A \propto e^{-K_A}$ of the subregion $A$
is a hermitean and positive semidefinite operator normalised by $\textrm{Tr}_{\!{}_A} \rho_A =1$.
The hermitean operator $K_A$ is the modular Hamiltonian (also known as entanglement Hamiltonian) of the region $A$
\cite{Haag:1992hx, takesaki-book} and its spectrum provides the entanglement entropy $S_A = -\, \textrm{Tr}_{\!{}_A}  (\rho_A  \log \rho_A)$.
The modular Hamiltonian $K_A$ leads to define the family of unitary operators 
$U(\tau)=e^{-\textrm{i} \tau K_A}$, parameterised by the modular parameter $\tau\in \RR$,
that generates the modular flow $\mathcal{O}(\tau) \equiv U(\tau) \,\mathcal{O}\,U(-\tau)$ of any operator $\mathcal{O}$ localised in $A$.
This modular flow describes the intrinsic internal dynamics induced by the reduced density matrix. 

It is important analytic expressions for the modular Hamiltonians in terms of the fundamental fields 
and for the corresponding modular flows.
The first seminal example, in generic spacetime dimensions, 
is the modular Hamiltonian of half space $x>0$ 
for a Lorentz invariant quantum field theory in its vacuum. 
This modular Hamiltonian, found by Bisognano and Wichmann \cite{Bisognano:1975ih,Bisognano:1976za},
is given by the boost generator in the $x$-direction.
In Conformal Field Theory, by combining the result of Bisognano and Wichmann with the conformal symmetry, 
some modular Hamiltonians can be written in explicit form \cite{Hislop:1981uh,Brunetti:1992zf,Casini:2011kv, Wong:2013gua, Cardy:2016fqc, Tonni:2017jom}.
All these modular Hamiltonians are local: they are written as an integral over $A$ of a local operator multiplied by a proper weight function. 

The first example of non-local modular Hamiltonian has been found by Casini and Huerta \cite{Casini:2009vk}
for the massless Dirac field in its ground state and on the infinite line, when the subsystem $A$ is the union of disjoint intervals,
by employing the lattice results for this operator obtained by Peschel \cite{Peschel:2003rdm} (see also the reviews \cite{Casini:2009sr, EislerPeschel:2009review}).
In \cite{Casini:2009vk} also the modular flow of the Dirac field has been found, 
while the two-point correlators along this evolution satisfying the Kubo-Martin-Schwinger (KMS) condition \cite{Haag:1992hx}
have been written in \cite{Longo:2009mn}.
Other modular Hamiltonians for the massless Dirac fermion containing non-local terms 
have been discussed in  \cite{Klich:2015ina, Blanco:2019xwi, Fries:2019ozf, Hollands:2019hje}.

In the examples of modular Hamiltonians mentioned above, 
the underlying system is invariant under spatial translations. 
The simplest way to break this symmetry in $1+1$ dimensions 
is to consider a quantum field theory on the half-line. 
For the massless Dirac field on the half line, the energy conservation imposed in any boundary conformal field theory \cite{Cardy:1984bb, Cardy:1986gw, Cardy:1989ir} 
allows only two kinds of boundary conditions \cite{Mintchev:2011mx, Smith:2019jnh}.
Correspondingly, two inequivalent models are defined: the vector phase and the axial phase.
Each phase is parameterised by an angle and characterised by specific conservation laws;
indeed, either the charge or the helicity is preserved but not both of them \cite{Liguori:1997vd}.
Instead, for the massless Dirac field on the line both these symmetries are conserved. 
In these two inequivalent phases, the modular Hamiltonians of an interval and the corresponding modular flows for the Dirac field have been studied in \cite{Mintchev:2020uom}.
These modular Hamiltonians contain bi-local terms and preserve the symmetry of the underlying phase.

The invariance under spatial translations on the line is broken also by introducing a point-like defect. 
A basic difference between boundaries and defects (see \cite{Andrei:2018die} for a recent review)
is that apart from reflection, the latter ones are able to transmit as well. 
In the theory of quantum transport \cite{landauer-57, landauer-70, buttiker-86, buttiker-88},
a defect is usually implemented by a one-body 
scattering matrix, which describes its interaction with the bulk particles. Such a scattering matrix can be introduced 
by adding to the bulk Hamiltonian an interaction term localised at the defect. This is for instance the conventional  
approach to the  Kondo effect \cite{Andrei-furuka-83,Affleck:1995ge, Ludwig:1994dy, Saleur:1998hq}. 
Another option, which works for point-like defects, is to impose 
boundary conditions on the bulk fields at the defect. This approach finds relevant applications to the 
transport in quantum wire junctions. The boundary conditions characterising the defect have an important 
physical impact. For quantum wires, where the universality in the bulk is described by a 
Luttinger liquid, the boundary conditions at the junction can give origin to rich phase diagrams \cite{Nayak:1999zz, Oshikawa:2005fh, Bellazzini:2008fu},
whose degree of universality has still not been fully understood. 
Instead, we recall in this respect that for one dimensional systems with a single boundary, conformal field theory 
provides a complete classification of the universality classes \cite{Cardy:1984bb, Cardy:1986gw, Cardy:1989ir}.

\begin{figure}[t!]
\vspace{-.3cm}
\hspace{-1cm}
%\begin{center}
\includegraphics[width=1.13\textwidth]{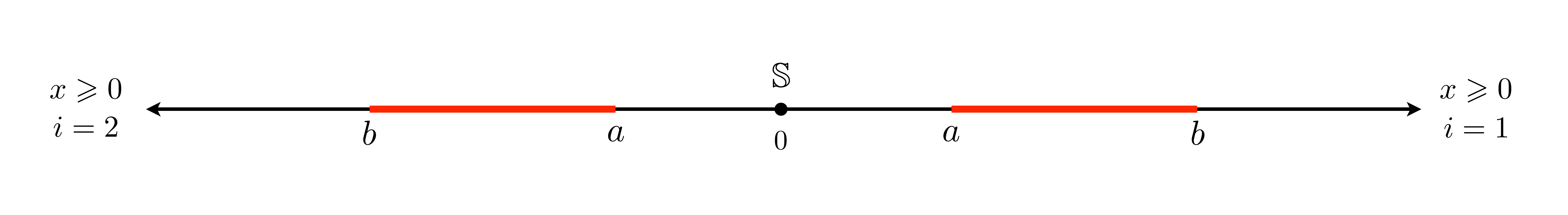}
% \end{center}
\vspace{-.9cm}
\caption{ 
The bipartition of the line mainly considered in this manuscript: 
Two disjoint equal intervals with length $b-a$ separated by a distance $2a$,
placed symmetrically with respect to a point-like defect 
described by the scattering matrix $\mathbb{S}$.
}
\label{fig-setup}
\end{figure}

The entanglement entropies, namely the entanglement entropy $S_A$ and the R\'enyi entropies
 $S_A^{(n)} \equiv \tfrac{1}{1-n} \log (\textrm{Tr} \rho_A^n)$ with $n\geqslant 2$,
have been studied in many models in the presence of a boundary or a defect
\cite{Calabrese:2004eu, AffleckLaFlorencie07, Sakai:2008tt, AffleckLaFlorencie09, 
EislerPeschel:2010def, Calabrese:2011ru, EislerPeschel:2012def, Saleur:2013pva, Ossipov14}.
Some local entanglement Hamiltonians in models with a boundary have been explored in 
\cite{Cardy:2016fqc, Tonni:2017jom, DiGiulio:2019cxv, Jensen:2013lxa, Mintchev:2020uom}.

In this manuscript we consider the massless Dirac fermion on the line in the presence of a point-like defect and in its ground state. 
The defect is characterised by a unitary scattering matrix $\SS$, 
which is determined by imposing the most general scale invariant boundary conditions without energy dissipations. 
In this setup, we derive the modular Hamiltonians of the union of two disjoint equal intervals 
at the same distance from the defect (see Fig.\,\ref{fig-setup}).
The associated modular flows are also investigated.
Our analysis heavily employs the results of \cite{Mintchev:2020uom} for the modular Hamiltonian of one interval on the half-line.

The outline of the manuscript is as follows. 
In Sec.\,\ref{sec_dirac_fermion} we discuss the massless Dirac fermion on the line in the presence of a defect,
introducing also the auxiliary fields.
These new fields are used in Sec.\,\ref{sec_eh} to write the modular Hamiltonians.
In Sec.\,\ref{sec_EE} we compute the entanglement entropies.
The modular flows and the correlation functions along them are discussed in Sec.\,\ref{sec_mod_flow} and in Sec.\,\ref{sec_correlators} respectively. 
In Sec.\,\ref{sec_limiting-regimes} we consider some limits of the spatial bipartition shown in Fig.\,\ref{fig-setup} where the modular Hamiltonians become local.
In Sec.\,\ref{space-time} we extend the modular Hamiltonians and their modular flows to a generic value of the physical time. 
The results are summarised and further discussed  in Sec.\,\ref{sec_conclusions}.

%\newpage
%%%%%%%%%%%%%%%%%%%%%%%%%%%%%%%%%%%%%%%%%%%

\section{Dirac fermions with a point-like defect on the line}
\label{sec_dirac_fermion}  

In this section we review the basic properties of the massless Dirac field with a point-like defect.
The defect, localised in the origin $x=0$ of the infinite line $\RR$ without loss of generality,
splits the line in two half-lines (edges). 
In order to treat these two edges in a symmetric way, we find it convenient to adopt the following coordinates 
\be
\big\{(x,i) \, :\, x>0,\, i=1,2\big\} 
\label{coordinates}
\ee
where $x$ indicates the distance from the defect and $i$ labels the edges, as shown in Fig.\,\ref{fig-setup}.

\subsection{General features}
\label{sec_generals}

The massless Dirac field $\psi (t,x,i)$ in the $i$-th half-line is the following doublet made by the two complex fields
\be
\label{psi-doublet}
\psi (t,x,i)=
\bigg(
\begin{array}{c}\psi_1(t,x,i) \\ \psi_2(t,x,i) \\ \end{array}
\bigg)
\ee 
whose dynamics in the coordinates (\ref{coordinates}) is described by the Dirac equation
\begin{equation}
\label{dirac-eom}
(\gamma_t \partial_t - \gamma_x \partial_x)\psi (t,x,i) = 0 
\;\;\qquad\;\;  x > 0
\end{equation} 
where 
\be
\gamma_t = 
\bigg(\begin{array}{cc}
0\; & 1 \\ 1 \; &  0
\end{array}  \bigg)
\;\;\qquad\;\;
\gamma_x 
= 
\bigg(\begin{array}{cc}
0 \; & -1\\ 1 \; &  0
\end{array}  \bigg)\,.
\end{equation}

The associated energy-momentum tensor is
\begin{eqnarray} 
\label{endens}
T_{tt}(t,x,i) &=& 
\frac{\ri}{2} \big[
(\prt_x \psi^*) \,\gamma_t \gamma_x \,\psi -
\psi^* \,\gamma_t \gamma_x \,(\prt_x \psi) 
\big](t,x,i) 
\\
\rule{0pt}{.7cm}
T_{xt}(t,x,i) &=&  \frac{\ri}{2} 
\big[
(\prt_t \psi^*) \, \gamma_t \gamma_x \, \psi -
\psi^*\, \gamma_t\gamma_x \,(\prt_t \psi) 
\big](t,x,i) 
\label{encurr}
\end{eqnarray} 
where ${}^*$ denotes Hermitean conjugation.

The bulk dynamics is invariant under both $U_{\textrm{\tiny v}}(1)$ vector 
and $U_{\textrm{\tiny a}}(1)$ axial phase transformations,
which are defined respectively by 
\begin{equation}
\psi_r (t,x,i) 
\; \longmapsto \;
\e^{\ri \, \theta_{\textrm{\tiny v}}} \,\psi_r(t,x,i) 
\;\;\qquad\;\; 
\psi_r (t,x,i) 
\; \longmapsto \;
\e^{\ri (-1)^s\, \theta_{\textrm{\tiny a}}}\, \psi_r(t,x,i)
\;\;\qquad\;\; 
\theta_{\textrm{\tiny v}}\, , \theta_{\textrm{\tiny a}} \in [0,2\pi)\,.
\label{va}
\end{equation}
The corresponding vector current $\{j_t, j_x\}$ and axial current $\{k_t, k_x\}$ are given by 
\begin{equation} 
j_t(t,x,i) = k_x(t,x,i) = [\psi^* \psi ](t,x,i) 
\;\;\qquad\;\; 
j_x(t,x,i) = k_t(t,x,i) = [\psi^* \gamma_t\gamma_x \psi] (t,x,i) 
\label{e2}
\end{equation}
which describe respectively the electric charge and helicity transport in the system.

The equation of motion (\ref{dirac-eom}) implies the local conservation laws in the bulk, namely
\begin{equation} 
\prt_t T_{tt}(t,x,i) - \prt_x T_{xt}(t,x,i) = 0 
\;\;\qquad\;\; 
x>0
\label{encons}
\end{equation}
and
\begin{equation}
\prt_t j_t(t,x) -\prt_x j_x(t,x,i) = 0
\;\;\qquad\;\;
\prt_t k_t(t,x) -\prt_x k_x(t,x,i) = 0 
\;\;\qquad\;\; 
x > 0 \,.
\label{currcons}
\end{equation} 

In order to determine the dynamics of the system on the line, 
we must fix the boundary conditions at $x=0$.
These boundary conditions define the defect. 

In this manuscript we consider the most general boundary conditions which ensure energy conservation.
This requirement is equivalent to impose \cite{Mintchev:2011mx} the validity of the  Kirchhoff rule 
\begin{equation}
\label{K}
\lim_{x\to 0} \, \sum_{i=1}^2 T_{xt}(t,x,i) = 0
\;\;\qquad\;\;  \forall t \in \RR \,.
\end{equation} 
In the presence of defects which allow both reflection and transmission,
the Kirchhoff rule (\ref{K}) is the counterpart of the Cardy condition \cite{Cardy:1984bb, Cardy:1986gw, Cardy:1989ir},
imposing the vanishing of the energy flow through a boundary. 
Using the explicit form of $T_{xt}$ in  (\ref{encurr}), we find that the condition (\ref{K}) can be satisfied in two ways:
either
\begin{equation}
\psi_1(t,0,i) = \sum_{j=1}^2 \SS^{\textrm{\tiny (v)}}_{ij}\, \psi_2(t,0,j) 
\label{bc1}
\end{equation}
or
\begin{equation}
\psi_1(t,0,i) =  \sum_{j=1}^2 \psi^*_2(t,0,j)\, \SS^{\textrm{\tiny (a)}*}_{ji}\, 
\label{bc2} 
\end{equation}
where $\SS^{\textrm{\tiny (v)}}$ and $\SS^{\textrm{\tiny (a)}}$ are generic $2\times 2$ unitary matrices
whose physical meaning is clarified below. 

Like on the half-line \cite{Mintchev:2011mx, Smith:2019jnh,Mintchev:2020uom},
the boundary conditions (\ref{bc1}) and (\ref{bc2}) are scale invariant
and determine the bulk internal symmetry group.  
In particular, the condition (\ref{bc1}) preserves $U_{\textrm{\tiny v}}(1)$, but breaks down 
$U_{\textrm{\tiny a}}(1)$. 
Instead, the condition (\ref{bc2}) preserves $U_{\textrm{\tiny a}}(1)$, breaking down $U_{\textrm{\tiny v}}(1)$. 
Thus, in the presence of an energy preserving defect,
one cannot keep both the $U_{\textrm{\tiny v}}(1)$ and $U_{\textrm{\tiny a}}(1)$ symmetries.
This means that  (\ref{bc1}) and (\ref{bc2}) define two inequivalent models,
called respectively vector phase and axial phase throughout the manuscript.

The components of the Dirac field satisfy the following anticommutation relations at equal time
\bea
\label{car1}
& &
\big[\, \psi_{r_1}(t,x_1,i_1)\, ,\, \psi^*_{r_2}(t,x_2,i_2)\,\big]_+ 
= \,
\delta_{r_1r_2}\, \delta_{i_1i_2}\, \delta (x_1-x_2)
\\
\rule{0pt}{.45cm}
\label{car2}
& &
\big[\, \psi_{r_1}(t,x_1,i_1)\, ,\, \psi_{r_2}(t,x_2,i_2)\, \big]_+
= \,
\big[\, \psi^*_{r_1}(t,x_1,i_1)\, ,\, \psi^*_{r_2}(t,x_2,i_2)\, \big]_+ =\,0
\eea

In order to construct the quantum fields satisfying the equation of motion 
(\ref{dirac-eom}), the equal time anticommutatots (\ref{car1})-(\ref{car2}) 
and the boundary conditions (\ref{bc1})-(\ref{bc2}), 
we introduce two CAR algebras ${\cal A}_+$ and ${\cal B}_+$ 
generated by 
\begin{equation} 
\{a_i(k),\, a_i^*(k)\, :\, k\geqslant 0,\, i=1,2\}\, ,\qquad \{b_i(k),\, b_i^*(k)\, :\, k\geqslant 0,\, i=1,2\}\, , 
\label{car}
\end{equation} 
which satisfy the canonical anti-commutation relation (CAR) 
and anti-commute each other.

In the vector phase, the components of the field can be decomposed as \cite{Mintchev:2011mx} 
\bea
\lambda_1(x+t,i) 
&=&
\int_0^{\infty} \frac{\rd k}{2\pi}\, \bigg[\,
a_i(k) \,\e^{-\ri k (x+t)} + \sum_{j=1}^2 \SS^{\textrm{\tiny (v)}}_{ij} b_j^*(k)\, \e^{\ri k (x+t)} 
\,\bigg]
\label{npsi1} 
\\
\lambda_2(x-t,i) 
&=& 
\int_0^{\infty} \frac{\rd k}{2\pi} \,\bigg[\,
\sum_{j=1}^2 \SS^{\textrm{\tiny (v)}*}_{ij} a_j(k) \,\e^{\ri k (x-t)} + b_i^*(k) \,\e^{-\ri k (x-t)}
\,\bigg]
\label{npsi2}
\eea
whereas in the axial phase the decomposition reads
\bea
\chi_1(x+t,i) 
&=&
\int_0^{\infty} \frac{\rd k}{2\pi} \, \bigg[\, \sum_{j=1}^2  b_j (k) \,\SS^{\textrm{\tiny (a)}*}_{ji}\e^{-\ri k (x+t)} + a^*_i(k) \,\e^{\ri k (x+t)}\, \bigg] 
\label{nchi1} 
\\
\chi_2(x-t,i) 
&=& \int_0^{\infty} \frac{\rd k}{2\pi} \, \bigg[\,\sum_{j=1}^2 \SS^{\textrm{\tiny (a)}*}_{ij}a_j(k) \e^{\ri k (x-t)} + b_i^*(k)\, \e^{-\ri k (x-t)} \,\bigg]\,.
\label{nchi2}
\eea

The two-point functions of these fields in the Fock representation of the CAR algebras ${\cal A}_+$ and ${\cal B}_+$
provide the basic ingredients to study the modular Hamiltonians. 

In the vector phase, one finds the following two-point vacuum expectation values
\bea
\langle \lambda_1(x_1+t_1,i_1) \, \lambda_1^*(x_2+t_2,i_2)\rangle 
&=&
\delta_{i_1i_2} \, C (t_{12}+x_{12})
\label{v11}
\\
\langle \lambda_2(x_1-t_1,i_1)\, \lambda_2^*(x_2-t_2,i_2)\rangle 
&=&
\delta_{i_1i_2} \, C(t_{12}-x_{12}) 
\label{v22}
\\
\langle \lambda_1(x_1+t_1,i_1)\,\lambda_2^*(x_2-t_2,i_2)\rangle 
&=&
 \SS^{\textrm{\tiny (v)}}_{i_1i_2}\, C (t_{12}+\tx_{12})
\label{v12}
\\
\langle \lambda_2(x_1-t_1,i_1)\,\lambda_1^*(x_2+t_2,i_2)\rangle 
&=&
\SS^{\textrm{\tiny (v)}*}_{i_1i_2} \, C(t_{12}-\tx_{12})
\label{v21}
\eea
where 
\begin{equation} 
C(\xi ) = \frac{1}{2\pi \ri (\xi - \ri \varepsilon)} = \frac{1}{2\pi \ri} \left [{\rm P.V.}\frac{1}{\xi} + \ri \pi \delta(\xi)\right ]  
\;\;\quad\;\; \varepsilon >0 
\label{v}
\end{equation}
and 
\begin{equation}
t_{12}=t_1-t_2  \;\;\qquad\;\;  x_{12}= x_1-x_2  \;\;\qquad\;\;  \tx_{21} = x_1+x_2\, . 
\label{not1}
\end{equation}
In the axial phase, a similar analysis leads to
\bea
\langle \chi_1^*(x_1+t_1,i_1)\, \chi_1(x_2+t_2,i_2)\rangle 
&=&
\delta_{i_1i_2} \, C (t_{12}+x_{12})
\label{a11}
\\
\langle \chi_2(x_1-t_1,i_1) \, \chi_2^*(x_2-t_2,i_2)\rangle 
&=&
\delta_{i_1i_2} \,C(t_{12}-x_{12}) 
\label{a22}
\\
\langle \chi_1^*(x_1+t_1,i_1)\, \chi_2^*(x_2-t_2,i_2)\rangle 
&=&
\SS^{\textrm{\tiny (a)}}_{i_1i_2} \,C (t_{12}+\tx_{12})
\label{a12}
\\
\langle \chi_2(x_1-t_1,i_1) \, \chi_1 (x_2-t_2,i_2)\rangle 
&=&
\SS^{\textrm{\tiny (a)}*}_{i_1i_2} \,C(t_{12}-\tx_{12})\,.
\label{a21}
\eea

In order to discuss both the phases in a unified way, let us  introduce the doublets 
\be
\label{lambdachi}
\psi (t,x,i) \equiv 
\bigg(\begin{array}{c} \psi_1(x+t,i) \\  \psi_2(x-t,i)  \end{array} \bigg)
\;\qquad\;
\lambda (t,x,i) \equiv 
\bigg(\begin{array}{c} \lambda_1(x+t,i) \\  \lambda_2(x-t,i)  \end{array} \bigg) 
\;\qquad\;
\chi (t,x,i) = 
\bigg(\begin{array}{c} \chi_1^*(x+t,i) \\  \chi_2(x-t,i)  \end{array} \bigg) 
\ee
and set the unifying notation
\be
\label{psiu}
\psi(t,x,i) 
\equiv 
\left\{\begin{array}{l}
\lambda (t,x,i) 
\\
\rule{0pt}{.5cm}
\chi (t,x,i) 
\end{array}
\right.
\;\;\;\;\qquad\;\;\;\;
\SS
\equiv 
\left\{\begin{array}{ll}
\SS^{\textrm{\tiny (v)}} \hspace{1.5cm}& \textrm{vector phase} 
\\
\rule{0pt}{.5cm}
\SS^{\textrm{\tiny (a)}} & \textrm{axial phase}\,.
\end{array}
\right.
\ee
In this notation the boundary conditions (\ref{bc1}) and (\ref{bc2}) at $x=0$ can be written as follows
\begin{equation}
\psi_1(t,i) = \sum_{j=1}^2 \SS_{ij}\, \psi_2(-t,j) \,.
\label{bc3}
\end{equation}

The $2\times 2$ matrix $\SS$ represents the unitary scattering matrix characterising the defect  \cite{Mintchev:2011mx}. 
In this respect, $|\SS_{11}|^2$ and $|\SS_{22}|^2$ describe the reflection probabilities from the defect, whereas 
$|\SS_{12}|^2$ and $|\SS_{21}|^2$ give the transmission probabilities between the two edges. The 
relative quantum scattering data are generated by 
the operators $\{a_i^*(k),\, b_i^*(k),\, a_i(k),\, b_i(k)\}$, which create and annihilate asymptotic particles 
with momentum $k$ in the $i$-th edge. 

By using the correlators (\ref{v11})-(\ref{v21}) and (\ref{a11})-(\ref{a21}),
one finds the following result for the density-density correlation function 
\bea 
\label{density-density}
& &
\langle j_t(t_1,x_1,i_1) \,j_t(t_2,x_2,i_2)\rangle 
\equiv 
\langle \,:\! \psi^* \psi \!:\!(t_1,x_1,i_1)\, :\!\psi^* \psi \!:\!(t_2,x_2,i_2)\,\rangle 
\\
& &
= 
\delta_{i_1i_2} 
\big[\,
C(t_{12} + x_{12})^2 + C(t_{12} - x_{12})^2 
\,\big] 
+ 
|\SS_{i_1i_2}|^2
 \big[\,
 C(t_{12} + \tx_{12})^2 + C(t_{12} - \tx_{12})^2 
 \,\big] 
\nonumber 
\eea
which nicely illustrates the impact of the $\SS$-matrix on the physical observables in the system.

The two extreme cases correspond to full reflection and full transmission,
defined respectively by 
\be 
\SS_{\textrm{\tiny ref}}
\equiv 
\bigg( \begin{array}{cc}1 \; & 0\\ 0 \; &  1 \\ \end{array} \bigg)
\;\;\;\;\qquad\;\;\;\;
\SS_{\textrm{\tiny tran}}
\equiv 
\bigg( \begin{array}{cc} 0 \; & 1\\  1 \; &  0 \\ \end{array} \bigg) 
\label{refltransm}
\ee 
In the case of full transmission, 
the components $\psi_{r}^{\textrm{\tiny line}}$ of the Dirac field  on the whole line $x\in \RR$ 
can be written in terms of the coordinates introduced in (\ref{coordinates}) as 
\be 
\label{psiline}
\psi_{1}^{\textrm{\tiny line}}(x+t) 
\equiv 
\left\{ \begin{array}{l}
\psi_1 (x+t,1) 
\\
\rule{0pt}{.5cm}
\psi_2 (-x-t,2)  
\end{array}\right.
\qquad 
\psi_{2}^{\textrm{\tiny line}}(x-t) 
\equiv 
\left\{ \begin{array}{ll}
\psi_2 (x-t,1) \hspace{1.2cm} &x>0   
\\
\rule{0pt}{.5cm}
\psi_1 (-x+t,2) &x<0  \,.
\end{array}\right.
\ee

We find it convenient to collect the two-point correlation functions at equal time 
$t_1=t_2\equiv 0$ into the following $4\times 4$ matrix
\be
\label{vcm0}
\left (
\begin{array}{cccc} 
\langle \psi_1 (x,1) \, \psi_1^*(y,1)\rangle \;\; & \langle \psi_1(x,1)\,\psi_2^*(y,1)\rangle \; \; 
& \langle \psi_1(x,1)\,\psi_1^*(y,2)\rangle\;\; & \langle \psi_1(x,1)\,\psi_2^*(y,2)\rangle 
\\ 
\rule{0pt}{.5cm}
\langle \psi_2 (x,1) \, \psi_1^*(y,1)\rangle \;\; & \langle \psi_2(x,1)\,\psi_2^*(y,1)\rangle \; \; 
& \langle \psi_2(x,1)\,\psi_1^*(y,2)\rangle\;\; & \langle \psi_2(x,1)\,\psi_2^*(y,2)\rangle 
\\
\rule{0pt}{.5cm}
\langle \psi_1 (x,2) \, \psi_1^*(y,1)\rangle \;\; & \langle \psi_1(x,2)\,\psi_2^*(y,1)\rangle \; \; 
& \langle \psi_1(x,2)\,\psi_1^*(y,2)\rangle\;\; & \langle \psi_1(x,2)\,\psi_2^*(y,2)\rangle 
\\
\rule{0pt}{.5cm}
\langle \psi_2 (x,2) \, \psi_1^*(y,1)\rangle \;\; & \langle \psi_2(x,2)\,\psi_2^*(y,1)\rangle \; \; 
& \langle \psi_2(x,2)\,\psi_1^*(y,2)\rangle\;\; & \langle \psi_2(x,2)\,\psi_2^*(y,2)\rangle 
\end{array} 
\right ) .
\ee
By using (\ref{v11})-(\ref{v21}), (\ref{a11})-(\ref{a21}) and the convention (\ref{psiu}), we find that this correlation matrix can be written as
\be
\label{vcm00}
\mathsf{C}(x,y; \SS)
=
\left (
\begin{array}{cccc} 
C(x-y) \;\; & \SS_{11}\,C(x+y) \; \; & 0 \;\; & \SS_{12}\,C(x+y) 
\\ 
\rule{0pt}{.5cm}
\SS^*_{11}\,C(-x-y) \;\; & C(-x+y) \; \; & \SS^*_{12}\,C(-x-y) \;\; & 0 
\\
\rule{0pt}{.5cm}
0 \;\; & \SS_{21}\,C(x+y) \; \; & C(x-y) \;\; & \SS_{22}\,C(x+y)  
\\
\rule{0pt}{.5cm}
\SS^*_{21}\,C(-x-y) \;\; & 0 \; \; & \SS^*_{22} \,C(-x-y) \;\; & C(-x+y) 
\end{array} 
\right ) .
\ee
This matrix, which is the basic input in the derivation of the modular Hamiltonian,
can be simplified by diagonalising $\SS$. 
This is achieved by introducing a new basis of auxiliary fields.

\subsection{Auxiliary fields basis}
\label{sec_aux_fields}

In order to deal with the correlation matrix (\ref{vcm0}), we find it convenient to introduce 
the unitary matrix $\U$ that diagonalises $\SS$, namely
\begin{equation} 
\U\, \SS\; \U^* 
= \bigg(\begin{array}{cc}\e^{\ri \alpha_1} \;& 0 \\ 0 \;&  \e^{\ri \alpha_2} \end{array} \!\bigg)
\;\;\qquad\;\; 
\U\; \U^* = \mathbb I\,.
\label{d}
\end{equation}
The unitary matrix $\U$  leads to define the auxiliary fields \cite{Calabrese:2011ru}
\be
\label{tildepsi}
\widetilde{\psi} (t,x,i) \equiv 
\Bigg(
\begin{array}{c} 
\widetilde{\psi}_1(x+t,i) 
\\ 
\rule{0pt}{.5cm}
\widetilde{\psi}_2(x-t,i)  \end{array} 
\Bigg) 
\ee
where 
\begin{equation}
\widetilde{\psi}_1 (x+t, i) \equiv \sum_{j=1}^2 \U_{ij}\, \psi_1 (x+t,j) 
\;\;\qquad \;\;
\widetilde{\psi}_2 (x-t, i) \equiv \sum_{j=1}^2 \U_{ij}\, \psi_2 (x-t,j)\,.
\label{Psi}
\end{equation}

We stress that ${\widetilde \psi}_1 (x+t, i)$ and ${\widetilde \psi}_2 (x-t, i)$ have unusual localisation:
they are given by a superposition of the local fields $\psi_1(x+t,j)$ and $\psi_2(x-t,j)$ respectively,
computed at the same time $t$ and distance 
$x$ from the defect but in different edges $j \in \{1,2\}$. 
However, the auxiliary fields provide a convenient 
basis to deal with the boundary conditions (\ref{bc3}) at $x=0$, 
which take the following simple diagonal form 
\be
{\widetilde \psi}_1 (t,i) = \e^{\ri \alpha_i} \,{\widetilde \psi}_2 (-t,i) 
\;\;\qquad \;\; 
i \in \{1,2\}\,.
\label{bc4}
\ee
Furthermore, the auxiliary fields obey the canonical equal-time relations (\ref{car1}) and (\ref{car2}).
This feature is essential in the derivation of the modular flow in Sec.\,\ref{sec_mod_flow}.
The physical observables and the corresponding correlation functions (see e.g. (\ref{density-density})) 
will be always expressed in terms of the physical doublet $\psi (t,x,i)$.

In the auxiliary field basis, the matrix of correlation functions reads
\be
%\widetilde{\mathsf{C}}(x,y)
%\equiv
\left (
\begin{array}{cccc} 
\langle \widetilde{\psi}_1 (x,1) \, \widetilde{\psi}_1^*(y,1)\rangle \;\; & \langle \widetilde{\psi}_1(x,1)\,\widetilde{\psi}_2^*(y,1)\rangle \; \; 
& \langle \widetilde{\psi}_1(x,1)\, \widetilde{\psi}_1^*(y,2)\rangle\;\; & \langle \widetilde{\psi}_1(x,1)\, \widetilde{\psi}_2^*(y,2)\rangle 
\\ 
\rule{0pt}{.5cm}
\langle \widetilde{\psi}_2 (x,1) \, \widetilde{\psi}_1^*(y,1)\rangle \;\; & \langle \widetilde{\psi}_2(x,1)\,\widetilde{\psi}_2^*(y,1)\rangle \; \; 
& \langle\widetilde{\psi}_2(x,1)\, \widetilde{\psi}_1^*(y,2)\rangle\;\; & \langle\widetilde{\psi}_2(x,1)\, \widetilde{\psi}_2^*(y,2)\rangle 
\\
\rule{0pt}{.5cm}
\langle \widetilde{\psi}_1 (x,2) \, \widetilde{\psi}_1^*(y,1)\rangle \;\; & \langle \widetilde{\psi}_1(x,2)\, \widetilde{\psi}_2^*(y,1)\rangle \; \; 
& \langle \widetilde{\psi}_1(x,2)\, \widetilde{\psi}_1^*(y,2)\rangle\;\; & \langle \widetilde{\psi}_1(x,2)\, \widetilde{\psi}_2^*(y,2)\rangle 
\\
\rule{0pt}{.5cm}
\langle \widetilde{\psi}_2 (x,2) \, \widetilde{\psi}_1^*(y,1)\rangle \;\; & \langle \widetilde{\psi}_2(x,2)\, \widetilde{\psi}_2^*(y,1)\rangle \; \; 
& \langle \widetilde{\psi}_2(x,2)\, \widetilde{\psi}_1^*(y,2)\rangle\;\; & \langle \widetilde{\psi}_2(x,2)\, \widetilde{\psi}_2^*(y,2)\rangle 
\end{array} 
\right ) .
\ee
A peculiar feature of this basis is that the correlation matrix becomes block-diagonal 
\be
\widetilde{\mathsf{C}}(x,y; \alpha_1, \alpha_2)
=
\boldsymbol{C}(x,y;\alpha_1)
\oplus
\boldsymbol{C}(x,y;\alpha_2)
\label{vcm1}
\ee

where 
\be
\label{cm2}
\boldsymbol{C}(x,y;\alpha ) 
\equiv
\Bigg(\,
\begin{array}{cc} 
C(x-y) \; & \e^{\ri \alpha} \,C(x+y)
\\ 
\rule{0pt}{.5cm}
\e^{-\ri \alpha}\, C(-x-y)  \; & C(-x+y)
\end{array} 
\Bigg)  .
\ee
This correlation matrix has been employed in \cite{Mintchev:2020uom} to determine the modular Hamiltonian 
of an interval in the half-line.

Using (\ref{d}) one finds that the matrices $\mathsf{C}(x,y;\SS)$ and $\widetilde{\mathsf{C}}(x,y;\alpha_1, \alpha_2)$ are unitarily 
equivalent 
\be 
\label{ueqv}
\boldsymbol{U}\, \mathsf{C}(x,y; \SS)\, \boldsymbol{U}^* = \widetilde{\mathsf{C}}(x,y;\alpha_1, \alpha_2)
\;\;\qquad\;\;
\boldsymbol{U}
\equiv  
\left (
\begin{array}{cccc} 
\U_{11} \;\; & 0 \; \; & \U_{12} \;\; & 0
\\ 
\rule{0pt}{.5cm}
0 \;\; & \U_{11} \; \; & 0 \;\; & \U_{12} 
\\
\rule{0pt}{.5cm}
\U_{21} \;\; & 0 \; \; & \U_{22} \;\; & 0  
\\
\rule{0pt}{.5cm}
0 \;\; & \U_{21} \; \; & 0 \;\; & \U_{22} 
\end{array} 
\right ) 
\ee 
where $\boldsymbol{U}$ is unitary because $\mathcal{U}$ is unitary. 

The equivalence (\ref{ueqv}) and the block-diagonal form (\ref{vcm1}) of $\widetilde{\mathsf{C}}(x,y; \alpha_1, \alpha_2)$ 
in terms of (\ref{cm2}) imply that the auxiliary field basis allows 
to study the system with a defect on the line by combining two independent half-line 
problems with the boundary conditions (\ref{bc4}). 
Indeed, the auxiliary fields 
$\widetilde{\psi}_{r_1}(x,i_1)$ and $\widetilde{\psi}_{r_2}(x,i_2)$ fully decouple for $i_1\not=i_2$. 
Accordingly, these fields have vanishing 
transmission probability in agreement with the diagonal form of the associated scattering matrix corresponding to (\ref{bc4}). 
As mentioned above, once the two independent half-line problems have been solved, 
their solutions must be combined to restore the local field $\psi(t,x,i)$-picture, 
which codifies the physical reflection-transmission properties of the defect defined by the original scattering matrix $\SS$. 

Our strategy in the following consists in first employing the results of \cite{Mintchev:2020uom} to write the modular Hamiltonians, 
the modular flows and the correlation functions for the auxiliary fields ${\widetilde \psi} (t,x,i)$,
then inverting (\ref{Psi}) to recover from them the corresponding quantities in the basis given by the physical fields $\psi(t,x,i)$.

We remark that the above setting has a natural generalisation 
to a space with the geometry of a star graph $\big\{(x,i) \, :\, x>0,\, i=1,...,n\big\}$ with $n$ edges \cite{Bellazzini:2006jb}. 
In that case, the defect is described by a $n\times n$ unitary scattering matrix  
and provides a physical model of a quantum wire junction \cite{Bellazzini:2008mn}.

%\newpage
%%%%%%%%%%%%%%%%%%%%%%%%%%%%%%%%%%%%%%%%%%%
\section{Modular Hamiltonians}
\label{sec_eh}

In this section we derive the modular Hamiltonians of the subregion $A_{\textrm{\tiny sym}}$ given by the union
of two disjoint equal intervals at the same distance from the defect (see Fig.\,\ref{fig-setup}),
by employing the results of \cite{Mintchev:2020uom}.

In the coordinates (\ref{coordinates}), the union of the two red segments in  Fig.\,\ref{fig-setup} reads
\be 
A_{\textrm{\tiny sym}}=\big\{(x,i) : 0< a\leqslant x \leqslant b\,,\, i=1,2 \big\}\,.
\label{domain}
\ee 
By introducing $A \equiv [a,b]$, the modular Hamiltonian is the following quadratic operator 
\cite{Peschel:2003rdm,Casini:2009sr, EislerPeschel:2009review}
\be
\label{K_A-phys-fields-0}
K_{A_{\textrm{\tiny sym}}} \,= 
\int_A \int_A
:\! \Psi^* (0,x) 
\, \log \! \big( \mathsf{C}_A(x,y; \SS)^{-1} - \mathbb{I}_4\,\big) \,
\Psi(0,y)\!:
\rd x  \,\rd y
\qquad
\Psi (0,x) \equiv 
\bigg(\begin{array}{c} \psi(0,x,1) \\  \psi(0,x,2)  \end{array} \bigg)
\ee
where $\mathsf{C}_A$ is the restriction of (\ref{vcm00}) to $A$,
the normal product in the CAR algebras ${\cal A}_+$ and ${\cal B}_+$ has been denoted by $:\cdots :$,
$\mathbb{I}_4$ is the $4\times 4$ identity matrix
and $\Psi$ has four components defined through (\ref{lambdachi}).

By employing the basis of auxiliary fields defined in (\ref{tildepsi}) and (\ref{Psi}),
together with (\ref{ueqv}), we find that the modular Hamiltonian (\ref{K_A-phys-fields-0}) can be written  in terms of the auxiliary fields as follows
\be
\label{K_A lambda}
K_{A_{\textrm{\tiny sym}}} \,= \sum_{i=1}^2
\int_A \int_A \!
:\! \tpsi^* (0,x,i) \,
\widetilde{\boldsymbol{H}}_A(x,y,i)\, 
\tpsi(0,y,i)\!:
\rd x  \,\rd y \,.
\ee
The kernel $\widetilde{{\boldsymbol H}}_A(x,y,i)$ is the $2\times 2$ matrix 
\be
\label{eh-matrix-peschel}
\widetilde{\boldsymbol{H}}_A(x,y,i) 
\equiv
\log \! \big(\boldsymbol{C}_A(x,y;\alpha_i)^{-1} - \mathbb{I}_2\big) 
\;\;\qquad\;\;
x,y \in A
\qquad
i\in \{1,2\}
\ee
where $\boldsymbol{C}_A$ is the reduced correlation functions matrix 
obtained by restricting (\ref{cm2}) to the interval $A$. 

The results of \cite{Mintchev:2020uom} imply that  
\be
\label{K_A-decomposition}
K_{A_{\textrm{\tiny sym}}}  \,= \, K_{A_{\textrm{\tiny sym}}}^{\textrm{\tiny loc}} + K_{A_{\textrm{\tiny sym}}}^{\textrm{\tiny bi-loc}} 
\ee 
where $K_{A_{\textrm{\tiny sym}}}^{\textrm{\tiny loc}}$ is a local operator, 
while $K_{A_{\textrm{\tiny sym}}}^{\textrm{\tiny bi-loc}} $ is a sum of bi-local operators. 

The local term in (\ref{K_A-decomposition}) reads
\be
\label{K_A-local-def}
K_{A_{\textrm{\tiny sym}}}^{\textrm{\tiny loc}} 
=
2\pi \sum_{i=1}^2
\int_a^b \!
\beta_{\textrm{\tiny loc}}(x) \, \widetilde{T}_{\textrm{\tiny loc}}(0,x,i)\, \rd x 
\,=\,
2\pi \sum_{i=1}^2
\int_a^b \!
\beta_{\textrm{\tiny loc}}(x) \, T_{tt}(0,x,i)\, \rd x 
\ee
where 
\be
\label{beta-loc-def}
\beta_{\textrm{\tiny loc}}(x) 
= 
\frac{(b^2 - x^2)\,(x^2 - a^2)}{2\,(b-a)\, (a\,b +x^2)}
\ee
and 
\be
\label{T00-lambda-def-tilde}
\widetilde{T}_{\textrm{\tiny loc}}(t,x,i) 
\equiv
\frac{\textrm{i}}{2}
:\! \! 
\Big[ \Big ((\partial_x \tpsi^\ast_1)\, \tpsi_1 - 
\tpsi^\ast_1\, (\partial_x \tpsi_1) \Big )(x+t,i)
- \Big((\partial_x \tpsi^\ast_2)\,  \tpsi_2 - \tpsi^\ast_2\, (\partial_x \tpsi_2)\Big)(x-t,i) 
\Big]\!\! : 
\,.
\ee
The last expression in (\ref{K_A-local-def}) is written in terms of the physical basis;
indeed,  from (\ref{Psi}) we have
\be 
\sum_{i=1}^2 \widetilde{T}_{\textrm{\tiny loc}}(t,x,i) = \sum_{i=1}^2 T_{tt}(t,x,i) 
\ee
where 
$T_{tt}(t,x,i)$ is the normal ordered version of the energy density (\ref{endens}), namely
\be
\label{T00-lambda-def}
T_{tt}(t,x,i) 
\equiv
\frac{\textrm{i}}{2}
:\! \!
\Big[ \Big ((\partial_x \psi^\ast_1)\, \psi_1 - 
\psi^\ast_1\, (\partial_x \psi_1) \Big )(x+t,i)
- \Big((\partial_x \psi^\ast_2)\,  \psi_2 - \psi^\ast_2\, (\partial_x \psi_2)\Big)(x-t,i) 
\Big]\!\! : 
\,.
\ee

As for the bi-local term in (\ref{K_A-decomposition}), from the results of \cite{Mintchev:2020uom}  for the auxiliary fields we find
\be
\label{K_A-bilocal-def}
K_{A_{\textrm{\tiny sym}}}^{\textrm{\tiny bi-loc}}  
=
2\pi \sum_{i=1}^2
\int_a^b \!
\beta_{\textrm{\tiny bi-loc}}(x) \, \widetilde{T}_{\textrm{\tiny bi-loc}}(0,x, \tilde{x}, i) \, dx
=
2\pi \sum_{i=1}^2
\int_a^b \!
\beta_{\textrm{\tiny bi-loc}}(x) \, T_{\textrm{\tiny bi-loc}}(0,x, \tilde{x}, i) \, dx
\;\;
\ee
where the weight function reads
\be
\label{beta-biloc-def}
\beta_{\textrm{\tiny bi-loc}}(x) 
\equiv
\frac{a\, b \, (b^2-x^2) \, (x^2-a^2)}{2\,(b-a)\, x\,(a\, b + x^2)^2} 
\ee
and
\be
\label{x-conjugate}
\tilde x \equiv \frac{ab}{x} 
\ee
with the bi-local operator defined as 
\bea
\label{T-bilocal-def}
\widetilde{T}_{\textrm{\tiny bi-loc}}(t,x, \tilde{x}, i) 
 &\equiv &
\frac{\textrm{i}}{2}\; 
\bigg\{ \,
e^{\textrm{i} \alpha_i}
\!:\!\!\Big[\, \tpsi^\ast_1(\tilde{x}+t,i) \,  \tpsi_2(x-t,i) + \tpsi^\ast_1(x+t,i) \,  \tpsi_2(\tilde{x}-t,i) \Big]\!\!: 
\\
& & \hspace{.8cm}
- \;
e^{-\textrm{i} \alpha_i}
\!:\!\! \Big[\, \tpsi^\ast_2({x}-t,i) \,  \tpsi_1(x+t,i) + \tpsi^\ast_2(x-t,i) \,  \tpsi_1(\tilde{x}+t,i) \Big] \!\!: \!
\bigg\}\,.
\nonumber
\eea
The bi-local term $K_{A_{\textrm{\tiny sym}}}^{\textrm{\tiny bi-loc}}  $  has been expressed in (\ref{K_A-bilocal-def}) also in terms of the physical fields by employing 
(\ref{Psi}), which give
\be
\label{sum-i-bilocal-term}
\sum_{i=1}^2 \widetilde{T}_{\textrm{\tiny bi-loc}}(t,x, \tilde{x}, i) 
=
\sum_{i=1}^2  T_{\textrm{\tiny bi-loc}}(t,x, \tilde{x}, i)
\ee
where we have introduced
\bea
\label{T-bilocal-defphys}
T_{\textrm{\tiny bi-loc}}(t,x, \tilde{x}, i)
 &= &
\frac{\textrm{i}}{2}\; \sum_{j=1}^2
\bigg\{ 
\!:\!\!\Big[\, \psi^\ast_1(\tilde{x}+t,i) \,\SS_{ij}\, \psi_2(x-t,j) + \psi^\ast_1(x+t,i) \, \SS_{ij} \, \psi_2(\tilde{x}-t,j) \Big]\!\!: 
 \\
& & \hspace{1.5cm}
- \;
\!:\!\! \Big[\, \psi^\ast_2(\tilde{x}-t,i)  \,\SS^*_{ij} \,\psi_1(x+t,j) + \psi^\ast_2(x-t,i)\,  \SS^*_{ij}\, \psi_1(\tilde{x}+t,j) \Big] \!\!: \!
\bigg\}\,.
 \hspace{.3cm}
\nonumber
\eea

We remark that, differently from the local term $K_{A_{\textrm{\tiny sym}}}^{\textrm{\tiny loc}}$, 
in the bi-local term $K_{A_{\textrm{\tiny sym}}}^{\textrm{\tiny bi-loc}} $ 
the left and right movers $\psi_1$ and $\psi_2$ are mixed through the scattering matrix $\SS$.

In the modular Hamiltonian of $A_{\textrm{\tiny sym}}$ on the line $\RR$ without defect \cite{Casini:2009vk, Longo:2009mn, Rehren:2012wa},
which corresponds to the limiting case of full transmission, 
the left-right mixing is absent and the counterpart of 
$K_{A_{\textrm{\tiny sym}}}^{\textrm{\tiny bi-loc}}$ depends only on 
two conjugate points, whose standard coordinates in $\RR$ are $x$ and $-\tilde x$.
Instead, in our case, the occurrence of the defect implies that,
in the standard coordinate on $\RR$, the bi-local term (\ref{K_A-bilocal-def})
involves two conjugate points for any given $x$: they are $\tilde{x}$ and $-\tilde{x}$
which can be interpreted respectively
as the reflected conjugate point (it also occurs for an interval in the half-line \cite{Mintchev:2020uom})
and as the transmitted conjugate point.  
These points play a distinguished role in the modular evolution described in Sec.\,\ref{sec_mod_flow}.

We find it worth comparing the modular Hamiltonians $K_{A_{\textrm{\tiny sym}}}$ in the two inequivalent phases. 
While the local term $T_{\textrm{\tiny loc}}(t,x,i)$ has the same form when expressed in terms 
of the fields $\lambda$ and $\chi$ given by (\ref{lambdachi}), 
crucial differences occur between the bi-local terms in the vector and in the axial phase. 
Indeed,  by using (\ref{psiu}) and (\ref{lambdachi}) in (\ref{T-bilocal-defphys}),
in the vector phase we have
\bea
T_{\textrm{\tiny bi-loc}}^{\textrm{\tiny vector}}(t,x, \tilde{x},i)
 &= &
\frac{\textrm{i}}{2}\; \sum_{j=1}^2
\bigg\{ 
\!:\!\!\Big[\, \lambda^\ast_1(\tilde{x}+t,i) \,\SS^{\textrm{\tiny (v)}}_{ij} \lambda_2(x-t,j) + \lambda^\ast_1(x+t,i)  
\,\SS^{\textrm{\tiny (v)}}_{ij}  \lambda_2(\tilde{x}-t,j) \Big]\!\!: 
\\
& & \hspace{1.6cm}
- \;
\!:\!\! \Big[\, \lambda^\ast_2(\tilde{x}-t,i) \, \SS^{\textrm{\tiny (v)}*}_{ij} \lambda_1(x+t,j) + \lambda^\ast_2(x-t,i)  
\,\SS^{\textrm{\tiny (v)}*}_{ij} \lambda_1(\tilde{x}+t,j) \Big] \!\!: \!
\bigg\}
\phantom{x}
\nonumber 
\label{T-bilocal-vect}
\eea
while in the axial phase the bi-local operator reads
\bea
T_{\textrm{\tiny bi-loc}}^{\textrm{\tiny axial}}(t,x, \tilde{x},i)
 &= &
\frac{\textrm{i}}{2}\; \sum_{j=1}^2
\bigg\{ 
\!:\!\!\Big[\, \chi_1(\tilde{x}+t,i) \,\SS^{\textrm{\tiny (a)}}_{ij} \chi_2(x-t,j) + \chi_2(x-t,i)\,  \SS^{\textrm{\tiny (a)}}_{ij}  \chi_1(\tilde{x}+t,j) \Big]\!\!: 
\\
& & \hspace{1.6cm}
- \;
\!:\!\! \Big[\, \chi^\ast_2(\tilde{x}-t,i) \, \SS^{\textrm{\tiny (a)}*}_{ij} \chi^\ast_1(x+t,j) + 
\chi^\ast_2(x-t,i) \, \SS^{\textrm{\tiny (a)}*}_{ij} \chi^\ast_1(\tilde{x}+t,j) \Big] \!\!: \!
\bigg\}\,.
\phantom{x}
\nonumber 
\label{T-bilocal-ax}
\eea
In the two inequivalent phase, 
the bi-local contribution has a different structure which respects their symmetry content. 
Indeed, the bi-local operator (\ref{T-bilocal-vect}) preserves the $U_{\textrm{\tiny v}}(1)$ symmetry, but breaks down $U_{\textrm{\tiny a}}(1)$; 
while the opposite holds for the bi-local operator (\ref{T-bilocal-ax}).

In the special cases of full reflection or full transmission, 
by employing (\ref{refltransm}) and (\ref{psiline}) into the above expressions,  
one recovers respectively either the modular Hamiltonian of an interval on the half-line \cite{Mintchev:2020uom} 
or modular Hamiltonian of two disjoint equal intervals on the line \cite{Casini:2009vk}.

%\newpage
%%%%%%%%%%%%%%%%%%%%%%%%%%%%%%%%%%%%%%%%%%%
\section{Entanglement entropies}
\label{sec_EE}

The entanglement entropies $S_A^{(n)} $ with $n \geqslant 1$ are obtained from
the moments of the reduced density matrix $\textrm{Tr}_{\!{}_A} \rho_A^n$ for integers $n \geqslant 2$.
These moments provide the R\'enyi entropies $S_A^{(n)} $ in a straightforward way
and the entanglement entropy $S_A$ through an analytic continuation of the integer parameter $n$ to real values as follows
\be
\label{renyi-def}
S_A^{(n)} \equiv \frac{1}{1-n}\, \log\!\big[ \textrm{Tr}_{\!{}_A} \rho_A^n\big]
\;\;\qquad\;\;
S_A \equiv \lim_{n \to 1} S_A^{(n)}
=
-\, \partial_n \big( \textrm{Tr}_{\!{}_A} \rho_A^n \big) \big|_{n=1}\,.
\ee

For the massless Dirac field, 
the entanglement entropies can be computed from the two-point correlators restricted to the subsystem $A$
as explained in \cite{EislerPeschel:2009review,Casini:2009sr}.

By employing the basis of the auxiliary fields, 
we can compute the entanglement entropies $S_A^{(n)}$ for the bipartition of the line shown in Fig.\,\ref{fig-setup},
when the entire system is in its ground state.
Since in the basis of the auxiliary fields the contributions corresponding to the two edges decouple, 
$S_A^{(n)}$ are obtained by summing these two contributions, which are equal.
The final result is twice the entanglement entropies 
of the bipartition given by an interval in the half-line, found in \cite{Mintchev:2020uom}, namely
\be
\label{ee-Asym-def}
S_{A_{\textrm{\tiny sym}}}^{(n)} 
= 
\frac{n+1}{3\, n}\,  \log\! \bigg[ \,\frac{2 \,\sqrt{a\,b}\,(b-a)}{(a+b) \,\epsilon} \, \bigg] + O(\epsilon) 
\ee
where $\epsilon>0$ is the ultraviolet cutoff and $a \gg \epsilon$.

The entanglement entropies of the bipartition of the line where the subsystem is an interval with the defect at its center,
i.e. $A = \{(x,i)\,:\, 0\leqslant x \leqslant a, \, i=1,2\}$,
which corresponds to the bipartition shown in Fig.\,\ref{fig-setup} in the limiting case where $a=0$,
can be studied in a similar way, by employing the results obtained in 
\cite{Mintchev:2020uom} for the interval adjacent to the boundary of the half-line.
Thus, for the entanglement entropies $S_{0,a}^{(n)}$ of an interval with the defect at its center we find
\be
\label{renyi-0}
S_{0,a}^{(n)} 
=
\frac{n+1}{6\, n}\,  \Big( \log(a/\epsilon)  + \log 2 \Big) + O(\epsilon)\,.
\ee

Combining (\ref{ee-Asym-def}) and (\ref{renyi-0}), we can construct the following ultraviolet finite quantity
\be
S_{0,a}^{(n)} + S_{0,b}^{(n)} - S_{A_{\textrm{\tiny sym}}}^{(n)} 
=
\frac{n+1}{3\, n}\,  \log \!\bigg(\frac{a+b}{b-a}\bigg)+ O(\epsilon) \,.
\ee

We remark that both the entanglement entropies 
(\ref{ee-Asym-def}) for the two disjoint equal intervals $A_{\textrm{\tiny sym}}$ and
(\ref{renyi-0}) for the interval with the defect in its center
are independent of the scattering matrix $\SS$.
This is due to the symmetry of these bipartition.
The independence of the scattering matrix 
for the entanglement entropies of an interval with the defect in its center
has been also observed for the non-relativistic fermion in \cite{Ossipov14}.
Instead, it is known that the entanglement entropies of bipartition which are not symmetric w.r.t. 
the position of the defect on the line 
depend on the parameters characterising the defect \cite{EislerPeschel:2010def, Calabrese:2011ru, EislerPeschel:2012def}.

The impurity entanglement entropies for the spatial configuration that we are considering 
are ultraviolet finite quantities obtained by subtracting $S_{A_{\textrm{\tiny sym}}}^{(n)} $ and 
the entanglement entropies of the same bipartition on the line without the defect \cite{AffleckLaFlorencie07}. 
In our case,  by using (\ref{ee-Asym-def}), one finds that the impurity entanglement entropies vanish.
This is due both to the symmetric choice of the bipartition and to the nature of the defect, which is scale invariant in our analysis. 
Examples where the impurity entanglement entropies are non trivial have been studied e.g. in \cite{AffleckLaFlorencie07,  AffleckLaFlorencie09, 
EislerPeschel:2010def, Calabrese:2011ru, EislerPeschel:2012def, Saleur:2013pva, Ossipov14}.

%\newpage
%%%%%%%%%%%%%%%%%%%%%%%%%%%%%%%%%%%%%%%%%%%
\section{Modular flows}
\label{sec_mod_flow}

In the following analysis we fix the time variable $t=0$ in (\ref{lambdachi}) and consider the modular evolutions 
of the fields $\psi_r (x,i)$ generated by the modular Hamiltonians introduced in Sec.\,\ref{sec_eh}.

The modular flow generated by $K_{A_{\textrm{\tiny sym}}}$ for the components of the massless Dira field is defined as
\be
\label{mod-evolution-lambda}
\psi_r(\tau,x,i) 
\,=\,
e^{-\textrm{i} \tau K_{A_{\textrm{\tiny sym}}}}  \, \psi_r(x,i)\, e^{\textrm{i} \tau K_{A_{\textrm{\tiny sym}}}}
\;\; \qquad \;\;
x\in A
\;\; \qquad \;\;
r\in \{1,2\}
\ee
and it can be determined by solving the initial value problem 
\be
\label{partial differential equation-mod-evolution}
\textrm{i}\,\frac{ d\psi_r(\tau,x,i)}{d\tau}  = \big[\,K_{A_{\textrm{\tiny sym}}} \,, \psi_r(\tau,x,i)\,\big]_{-}
\;\; \qquad \;\;
\psi_r(\tau=0,x,i) = \psi_r(x,i)\,
\ee

This modular flow can be obtained by first finding the modular flow of the auxiliary fields introduced in Sec.\,\ref{sec_aux_fields},
which is given by the solution of 
\be
\label{auxmodeq}
\textrm{i}\,\frac{ d\tpsi_r(\tau,x,i)}{d\tau}  = \big[\,K_{A_{\textrm{\tiny sym}}} \,, \tpsi_r(\tau,x,i)\,\big]_{-}
\;\; \qquad \;\;
\tpsi_r(\tau=0,x,i) = \tpsi_r(x,i)
\ee
and then writing the result in the physical basis. 

The commutator in (\ref{auxmodeq}) can be computed by employing 
the expression of the modular Hamiltonians in terms of the auxiliary fields and the fact that
the equal-time canonical anticommutation relations hold for the auxiliary fields.
By introducing 
\be
\label{Lambda-doublet-def}
\tPsi(\tau,x,i)
\equiv
\bigg(
\begin{array}{c}
\tpsi_1(\tau,x,i) \\  \rule{0pt}{.3cm} \tpsi_2(\tau,\tilde{x},i) 
\end{array}  \bigg) 
\ee
after some algebra we find
\be
\label{mod-flow-4-matrix-bdy}
\frac{d}{d\tau}
\bigg(
\begin{array}{c}
\tPsi(\tau, x,i) \\ \tPsi(\tau, \tilde{x},i) 
\end{array}  
\bigg) 
=\, 
2\pi \,\big[\, \boldsymbol{B}(x,i) \oplus \boldsymbol{B}(\tilde{x},i)\,\big] 
\bigg(
\begin{array}{c}
\tPsi(\tau, x,i) \\ \tPsi(\tau, \tilde{x},i) 
\end{array}  
\bigg) \qquad i=1,2\,.
\ee
The $4\times 4$ block diagonal matrix within the square brackets in the r.h.s.
is expressed in terms of the $2\times 2$ matrix 
\be
\boldsymbol{B}(x,i)
\equiv
\bigg(
\begin{array}{cc}
B_{\textrm{\tiny loc}}(x) & - \,e^{\textrm{i} \alpha_i}  \beta_{\textrm{\tiny bi-loc}}(x) 
\\
e^{-\textrm{i} \alpha_i}  \beta_{\textrm{\tiny bi-loc}}(\tilde{x}) & - \,B_{\textrm{\tiny loc}}(\tilde{x}) 
\end{array} 
\bigg) 
\ee
where $B_{\textrm{\tiny loc}}(y)$ is the following differential operator 
\be
\label{B-func-def}
B_{\textrm{\tiny loc}}(y) 
\equiv
\beta_{\textrm{\tiny loc}}(y)\, \partial_y + \frac{1}{2}\, \partial_y \beta_{\textrm{\tiny loc}}(y)
\ee
and the weight functions $\beta_{\textrm{\tiny loc}}(y)$ and $\beta_{\textrm{\tiny bi-loc}}(y)$ are given by (\ref{beta-loc-def}) and (\ref{beta-biloc-def}) respectively.

At this point we recognise that the equations in (\ref{mod-flow-4-matrix-bdy}) for the auxiliary fields 
are two independent copies  labelled by $i \in \{1,2\}$ 
of the modular flow equations for an interval on the half-line studied in \cite{Mintchev:2020uom}.
By employing the results of \cite{Mintchev:2020uom}, we find that the modular flow for the auxiliary fields with $x\in (a,b)$ reads
\be
\label{psi-mod-flow-ab0}
\left\{ \begin{array}{l}
\displaystyle
\tpsi_1(\tau,x,i)
=
\left[
P(\xi;  x) 
\left(
 \big( a \,b + x \,\xi \big) \,\tpsi_1(\xi,i) 
- \frac{a\,b}{\xi}\; e^{\textrm{i}\alpha_i} \big(\xi - x\big)\, \tpsi_2(ab/\xi,i)
\right)
\right]
\! \bigg|_{\xi = \xi(\tau,x)}
\\
\rule{0pt}{.9cm}
\displaystyle
\tpsi_2(\tau,x,i)
=
\left[
P(\xi;  x) 
\left(
 \big( a \,b + x \,\xi \big) \,\tpsi_2(\xi,i) 
- \frac{a\,b}{\xi}\; e^{-\textrm{i}\alpha_i} \big(\xi - x\big)\, \tpsi_1(ab/\xi,i)
\right)\right]
\! \bigg|_{\xi = \xi(-\tau,x)}
\end{array}
\right.
\ee
where
\be
P(\xi;  x)
\equiv
\sqrt{\frac{\beta_{\textrm{\tiny loc}}(\xi)}{\beta_{\textrm{\tiny loc}}(x)\, (a \,b + x^2)(a\,b+\xi^2) }}
\ee
and 
\be
\label{xi-def}
\xi(\tau,x) 
\,=\,
\frac{ (b-a)\big(e^{2\pi \tau + w(x)}-1\big) 
+
\sqrt{(b-a)^2\big(e^{2\pi \tau + w(x)}-1\big)^2 + 4 ab \,\big(e^{2\pi \tau + w(x)}+1\big)^2}
}{
2\,\big(1+e^{2\pi \tau + w(x)}\big)}
\ee
with 
\be 
w(x) = \log \left [\frac{(x+b)(x-a)}{(x+a)(b-x)}\right ] .
\label{w}
\ee
The function $\xi(\tau, x)$ describes the modular evolution of any point $x \in A$ and satisfies
\be
\label{x-tilde-xi-tilde}
\xi(0,x) =x \qquad 
\xi(\tau,x) \in (a,b)\qquad  
\textrm{and} \qquad
\xi(-\tau, \tilde{x}) = \frac{a\, b}{\xi(\tau,x)} \equiv \tilde{\xi}(\tau,x)\,.
%\equiv \tilde{\xi}(\tau, x) 
\ee
This guarantees that the solution (\ref{psi-mod-flow-ab0}) fulfils  the initial condition $\tpsi_r(0,x,i)= \tpsi_r(x,i)$,
where $\tpsi_r(x,i)$ provide the assigned initial configuration for the auxiliary fields on the line.

Combining the solution (\ref{psi-mod-flow-ab0}) for the auxiliary fields with  (\ref{d}), (\ref{tildepsi}) and (\ref{Psi}),
we obtain the modular flow of the physical fields
\be
\label{psi-mod-flow-ab}
\left\{ \begin{array}{l}
\displaystyle
\psi_1(\tau,x,i)
=
\Bigg[
P(\xi;  x) 
\Bigg(
 \big( a \,b + x \,\xi \big) \,\psi_1(\xi,i) 
- \frac{a\,b}{\xi} \big(\xi - x\big) \sum_{j=1}^2 \SS_{ij}\, \psi_2(ab/\xi,j)
\Bigg)
\Bigg]
\! \Bigg|_{\xi = \xi(\tau,x)}
\\
\rule{0pt}{.9cm}
\displaystyle
\psi_2(\tau,x,i)
=
\Bigg[
P(\xi;  x) 
\Bigg(
 \big( a \,b + x \,\xi \big) \,\psi_2(\xi,i) 
- \frac{a\,b}{\xi} \big(\xi - x\big) \sum_{j=1}^2 \SS^*_{ij}\, \psi_1(ab/\xi,j)
\Bigg) \Bigg]
\! \Bigg|_{\xi = \xi(-\tau,x)}
\end{array}
\right.
\ee 
which is one of the main results of this paper. 
Considering e.g. the first expression in (\ref{psi-mod-flow-ab}),
we observe that the modular evolution of $\psi_1(\tau,x,i)$ is determined not only 
by the modular evolution of the initial data for $\psi_1(x,i)$ along $(\xi (\tau, x),i)$,
but also by the modular evolution  of the initial data for $\psi_2(x,i)$ along 
the two conjugate trajectories $(\tilde \xi (\tau,x),1)$ and $(\tilde \xi (\tau,x),2)$. 
From the second expression in (\ref{psi-mod-flow-ab}), we have that a similar consideration holds for $\psi_2(\tau,x,i)$, 
with $\xi (\tau,x)$ replaced by $\xi (-\tau,x)$.
This is a consequence of the fact that the scattering matrix $\SS$,
which occurs explicitly in the flow, allows both reflection and transmission.

\begin{figure}[t!]
\vspace{-.5cm}
\hspace{-1cm}
%\begin{center}
\includegraphics[width=1.12\textwidth]{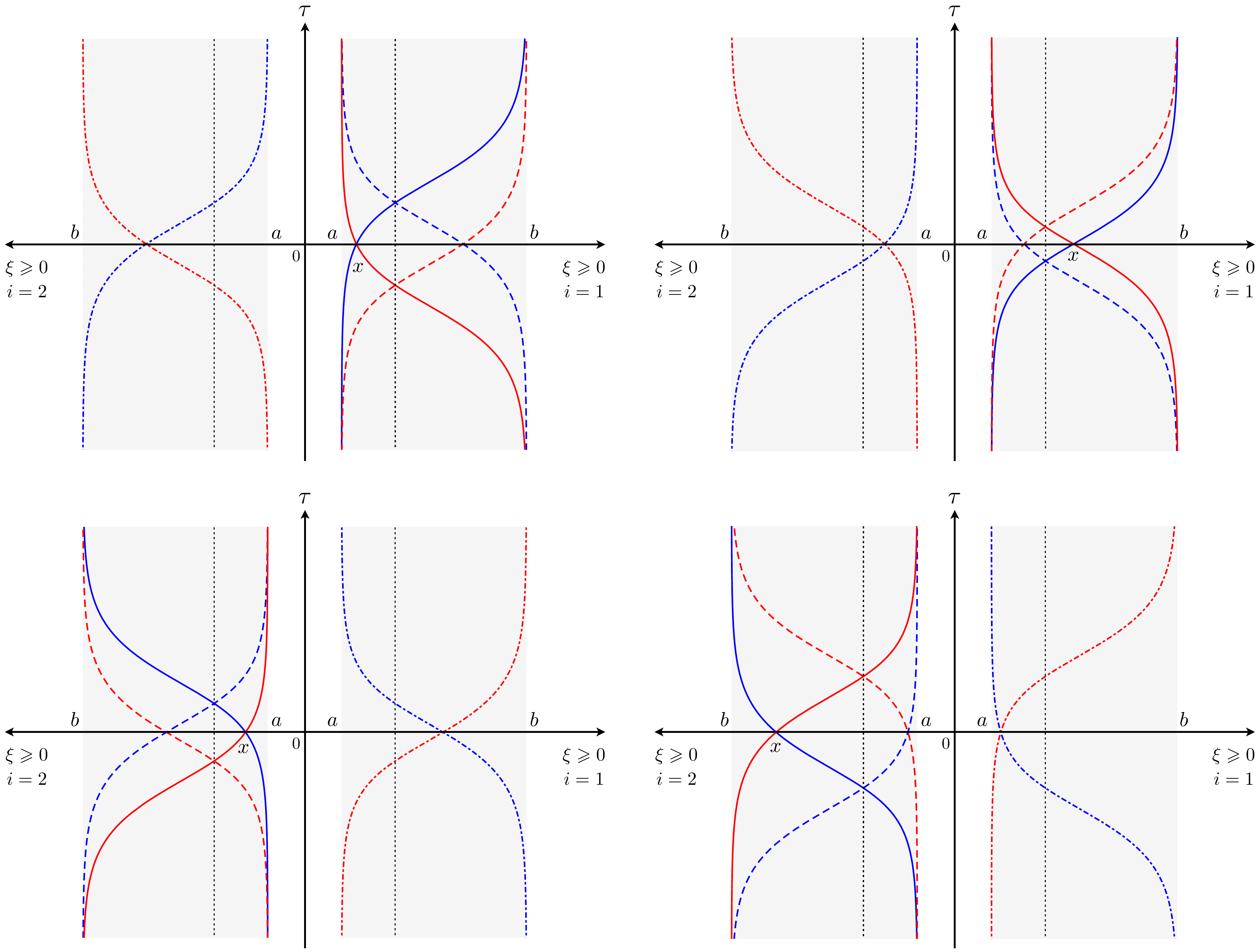}
% \end{center}
\vspace{-.2cm}
\caption{ 
Modular evolutions of the arguments of the fields mixed by the modular flow in the r.h.s's of (\ref{psi-mod-flow-ab}),
for a point $(x,i)$ (solid lines) and its conjugate points $(\tilde{x},i)$ (dashed lines) and $(\tilde{x}, j \neq i)$ (dashed-dotted lines), at $\tau=0$.
The blue and the red curves correspond to the first and to the second equation in (\ref{psi-mod-flow-ab}) respectively.
The vertical dotted lines identify the point $\sqrt{ab}$ in the two edges. 
Top panels: $i=1$ with either $x<\sqrt{ab}$ (left) or $x>\sqrt{ab}$ (right).
Bottom panels: $i=2$ with either $x<\sqrt{ab}$ (left) or $x>\sqrt{ab}$ (right).
}
\label{fig-xi}
\end{figure}

In Fig.\,\ref{fig-xi} we show
the modular evolution of the arguments of the fields mixed by the modular flow (\ref{psi-mod-flow-ab}), 
for four different choices of the initial point in $A_{\textrm{\tiny sym}}$ at $\tau =0$.
When $x\neq \sqrt{ab}$, three distinct curves occur and the ones in the same edge
intersect at either $|\tau_0|$ or $-|\tau_0|$, 
where $2\pi \tau_0 = | w(\sqrt{ab}\,) -w(x) | = | w(x) |$ because  $w(\sqrt{ab}\,) =0$.

For full reflection or full transmission, from (\ref{refltransm}) and (\ref{psiline}),
the sum in the r.h.s. of (\ref{psi-mod-flow-ab}) involves only one term and, 
accordingly, we recover the modular flow of the massless Dirac field 
either for an interval on the half-line \cite{Mintchev:2020uom} 
or for $A_{\textrm{\tiny sym}}$ on the whole line \cite{Casini:2009vk} respectively.

We find it worth reporting the explicit expressions of the modular flows in the two phases.

By using (\ref{lambdachi}) and (\ref{psiu}) in (\ref{psi-mod-flow-ab}),
in the vector phase we have
\be
\label{psi-mod-flow-vector}
\left\{ \begin{array}{l}
\displaystyle
\lambda_1(\tau,x,i) = 
\Bigg[
P(\xi;  x) 
\Bigg(
 \big( a \,b + x \,\xi \big) \,\lambda_1(\xi,i) 
- \frac{a\,b}{\xi} \big(\xi - x\big) \sum_{j=1}^2 \SS^{\textrm{\tiny (v)}}_{ij}\, \lambda_2(ab/\xi,j)
\Bigg)
\Bigg]
\! \Bigg|_{\xi = \xi(\tau,x)}
\\
\rule{0pt}{.9cm}
\displaystyle
\lambda_2(\tau,x,i)
=
\Bigg[
P(\xi;  x) 
\Bigg(
 \big( a \,b + x \,\xi \big) \,\lambda_2(\xi,i) 
- \frac{a\,b}{\xi} \big(\xi - x\big) \sum_{j=1}^2 \SS^{\textrm{\tiny (v)}*}_{ij}\, \lambda_1(ab/\xi,j)
\Bigg)
\Bigg]
\! \Bigg|_{\xi = \xi(-\tau,x)}
\end{array}
\right.
\ee
while the modular flow in the axial phase reads
\be
\label{psi-mod-flow-axial}
\left\{ \begin{array}{l}
\displaystyle
\chi_1(\tau,x,i)
=
\Bigg[
P(\xi;  x) 
\Bigg(
 \big( a \,b + x \,\xi \big) \,\chi_1(\xi,i) 
- \frac{a\,b}{\xi} \big(\xi - x\big)  \sum_{j=1}^2 \chi_2(ab/\xi,j)\, \SS^{\textrm{\tiny (a)}*}_{ji}
\Bigg)
\Bigg]
\! \Bigg|_{\xi = \xi(\tau,x)}
\\
\rule{0pt}{.9cm}
\displaystyle
\chi_2(\tau,x,i)
=
\Bigg[
P(\xi;  x) 
\Bigg(
 \big( a \,b + x \,\xi \big) \,\chi_2(\xi,i) 
- \frac{a\,b}{\xi} \big(\xi - x\big)  \sum_{j=1}^2 \SS^{\textrm{\tiny (a)}*}_{ij}\, \chi_1(ab/\xi,j)
\Bigg)
\Bigg]
\! \Bigg|_{\xi = \xi(-\tau,x)}
\end{array}
\right.\,.
\ee

Each modular flow preserves the symmetry characterising its modular Hamiltonian.
Indeed, in the vector phase, the modular flow (\ref{psi-mod-flow-vector}) preserves the $U_{\textrm{\tiny v}}(1)$ symmetry and breaks the $U_{\textrm{\tiny a}}(1)$ symmetry,
while, in the axial phase, the modular flow (\ref{psi-mod-flow-axial}) preserves the $U_{\textrm{\tiny a}}(1)$ symmetry and breaks the $U_{\textrm{\tiny v}}(1)$ symmetry.

%\newpage
%%%%%%%%%%%%%%%%%%%%%%%%%%%%%%%%%%%%%%%%%%%
\section{Correlation functions along the modular flows}
\label{sec_correlators}

The modular evolutions (\ref{psi-mod-flow-vector}) and (\ref{psi-mod-flow-axial}) 
provide the corresponding correlation functions in the corresponding phase, 
which describe the quantum fluctuations along the modular evolution parameterised by $\tau$.

The initial data involved in (\ref{psi-mod-flow-vector}) and (\ref{psi-mod-flow-axial}) 
are expressed via (\ref{npsi1})-(\ref{nchi2}) in terms of the generators of the CAR algebras ${\cal A}_+$ and ${\cal B}_+$. 
Adopting the Fock representation for these algebras, one derives the correlation functions in the presence of the defect 
in closed and explicit form. 
Similar calculations have been done for the massless Dirac field in the ground state, 
when the subsystem is the union of disjoint intervals in the line \cite{Longo:2009mn,Hollands:2019hje}
or an interval in the half-line \cite{Mintchev:2020uom}.

Interestingly, all the correlation functions along the modular flow can be written through the distribution 
\be
\label{Wfunc}
W(\tau; x, y) 
\equiv
\frac{e^{w(x)} - e^{w(y)}}{2\pi \textrm{i} (x - y)}\;
\frac{1}{e^{w(x)+\pi \tau} - e^{w(y)-\pi \tau} - \textrm{i} \varepsilon} 
\ee
where $w(x)$ is the function defined in (\ref{w}). 
In the limit $\varepsilon \to 0$, 
the expression in (\ref{Wfunc}) satisfies
\be
\label{Widentity}
W(\tau \pm \textrm{i} \,; x, y) 
=
W(-\tau ; y, x) 
=
\overline{W(\tau ; x, y) }
\ee
where the overline denotes the complex conjugation. 

In the vector phase, the non-vanishing two-point functions take the form 
\bea
\label{corr-11-mod}
\langle \lambda_1(\tau_1,x_1,i_1)\,\lambda_1^*(\tau_2,x_2,i_2)\rangle 
&=&
\langle \lambda^\ast_1(\tau_1,x_1,i_1)\,\lambda_1(\tau_2,x_2,i_2)\rangle 
=
\delta_{i_ii_2}\, W(\tau_{12};x_1,x_2) 
\\
\label{corr-22-mod}
\langle \lambda_2(\tau_1,x_1,i_1)\,\lambda_2^*(\tau_2,x_2,i_2)\rangle 
&=&
\langle \lambda^\ast_2(\tau_1,x_1,i_1)\,\lambda_2(\tau_2,x_2,i_2)\rangle 
=  
\delta_{i_1i_2}\, W(\tau_{12};x_2,x_1) 
\\
\label{corr-mixed1-mod}
\langle \lambda_1(\tau_1,x_1,i_1)\,\lambda_2^*(\tau_2,x_2,i_2)\rangle 
&=&
\overline{\langle \lambda_2(\tau_2,x_2,i_2)\,\lambda_1^*(\tau_1,x_1,i_1)\rangle}
= 
\SS^{\textrm{\tiny (v)}}_{i_1i_2}\, W(\tau_{12};x_1,-x_2)
\\
\label{corr-mixed2-mod}
\langle \lambda_1^\ast(\tau_1,x_1,i_1)\,\lambda_2(\tau_2,x_2,i_2)\rangle 
&=&
\overline{ \langle \lambda_2^\ast(\tau_2,x_2,i_2)\,\lambda_1(\tau_1,x_1,i_1)\rangle }
=
\SS^{\textrm{\tiny (v)}*}_{i_1i_2} \;
W(\tau_{12};x_1,-x_2)
\hspace{.7cm}
\eea
with 
\be 
\tau_{12} \equiv \tau_1 - \tau_2 \,.
\ee 
 
Using the first identity in (\ref{Widentity}), one verifies that the correlation functions 
(\ref{corr-11-mod})-(\ref{corr-mixed2-mod}) satisfy the Kubo-Martin-Schwinger (KMS) condition 
\bea
\label{KMS1}
\langle \,\lambda_{r_1}(\tau_1,x_1)\, \lambda_{r_2}^*(\tau_2+\tau+\textrm{i},x_2)\, \rangle
&=&
\langle \,\lambda_{r_2}^*(\tau_2+\tau,x_2)\,\lambda_{r_1}(\tau_1,x_1)\,  \rangle
\\
\label{KMS2}
\langle \,\lambda^\ast_{r_1}(\tau_1,x_1)\, \lambda_{r_2}(\tau_2+\tau+\textrm{i},x_2)\, \rangle
&=&
\langle \,\lambda_{r_2}(\tau_2+\tau,x_2)\,\lambda^\ast_{r_1}(\tau_1,x_1)\,  \rangle
\eea
where $r_1, r_2 \in \{1,2\}$. The validity of (\ref{KMS1}) and (\ref{KMS2}) is a crucial feature 
of the modular group (see Theorem 1.2 in chapter VIII of \cite{takesaki-book}),
hence it also provides a valuable consistency check of our results.

Besides the KMS conditions, the correlation functions (\ref{corr-11-mod})-(\ref{corr-mixed2-mod}) 
satisfy also the modular equations of motion following from (\ref{partial differential equation-mod-evolution}). 
An illustrative example is
\be
\label{modeqdef}
\hspace{-.25cm}
\left[
\frac{1}{2\pi}\,\partial_{\tau_1} - B_{\textrm{\tiny loc}}(x_1) \right]\!
\langle \lambda_1(\tau_1,x_1,i_1)\,\lambda_1^*(\tau_2,x_2,i_2)\rangle 
=
- \,\beta_{\textrm{\tiny bi-loc}}(x_1)  
\sum_{j=1}^2 \SS^{\textrm{\tiny (v)}}_{i_1j} \, \langle \lambda_2(\tau_1,{\tilde x}_1,j)\,\lambda_1^*(\tau_2,x_2,i_2)\rangle 
\ee
with $x_1\not=x_2$. 
From (\ref{corr-11-mod}) and (\ref{corr-mixed1-mod}), we have that (\ref{modeqdef}) is equivalent to 
\be
\label{Weqdef}
\left[\,\frac{1}{2\pi}\,\partial_{\tau} - B_{\textrm{\tiny loc}}(x) \right ]W(\tau;x,y) 
+
 \beta_{\textrm{\tiny bi-loc}}(x) \, W(\tau;-{\tilde x},y) 
\,=\,
0  
\ee
whose validity in the limit $\varepsilon \to 0$ follows directly from the definition (\ref{Wfunc}).

The correlation functions (\ref{corr-11-mod})-(\ref{corr-mixed2-mod}) in the vector phase have a direct physical application 
to the electric an helical transport across the defect.
Indeed, they lead to the density-density correlators
\bea 
\label{cc1}
& &
\langle j_t(\tau_1,x_1,i_1)\, j_t(\tau_2,x_2,i_2)\rangle = \langle k_x(\tau_1,x_1,i_1)\, k_x(\tau_2,x_2,i_2)\rangle 
\\
& &
= \delta_{i_1i_2} \Big[
W(\tau_{12};x_1,x_2)^2 + W(\tau_{12};x_2,x_1)^2 
\Big] 
+  \big|\SS^{\textrm{\tiny (v)}}_{i_1i_2}\big|^2 
\Big[W(\tau_{12};x_1,-x_2)^2 + W(\tau_{12};-x_1,x_2)^2 \Big] 
\nonumber
\eea 
and to the current-current correlators
\bea 
\label{cc2}
& &
\langle j_x(\tau_1,x_1,i_1)\, j_x(\tau_2,x_2,i_2)\rangle = \langle k_t(\tau_1,x_1,i_1)\, k_t(\tau_2,x_2,i_2)\rangle 
\\
& &
= \delta_{i_1i_2} 
\Big[ 
W(\tau_{12};x_1,x_2)^2 + W(\tau_{12};x_2,x_1)^2 
\Big] 
-  \big|\SS^{\textrm{\tiny (v)}}_{i_1i_2}\big|^2 
\Big[W(\tau_{12};x_1,-x_2)^2 + W(\tau_{12};-x_1,x_2)^2 \Big] 
\nonumber
\eea 
which depend explicitly on the reflection and transmission probabilities $|\SS^{\textrm{\tiny (v)}}_{i_1i_2}|^2$. 
In agreement with conservation of the $U_{\textrm{\tiny v}}(1)$ symmetry in the vector phase, 
the correlator (\ref{cc2}) satisfies the Kirchhoff law at the defect in $x_1=0$; indeed
\bea 
\label{Kirchhoff3}
& &
\sum_{i_1=1}^2 \langle j_x(\tau_1,0,i_1)\, j_x(\tau_2,x_2,i_2)\rangle = 
\\
& &
=
\sum_{i_1=1}^2 \Big[
\left (\delta_{i_1i_2} -|\SS^{\textrm{\tiny (v)}}_{i_1i_2}|^2 \right ) W(\tau_{12};0,x_2) + 
\delta_{i_1i_2} W(\tau_{12};x_2,0)  - |\SS^{\textrm{\tiny (v)}}_{i_1i_2}|^2 \,W(\tau_{12};0,-x_2) 
\Big] 
\nonumber 
\\
& &
=
\sum_{i_1=1}^2 \left (\delta_{i_1i_2} -|\SS^{\textrm{\tiny (v)}}_{i_1i_2}|^2 \right ) \Big[W(\tau_{12};0,x_2) + W(\tau_{12};x_2,0) \Big]  =0 
\nonumber 
\eea
where we have employed the unitarity of $\SS^{\textrm{\tiny (v)}}$ and the identity 
\be 
W(\tau; x,0) = W(\tau;0,-x)\,.
\ee 
Notice that, instead, the helical current violates the Kirchhoff law 
in the vector phase 
\be
\sum_{i_1=1}^2 \langle k_x(\tau_1,0,i_1)\, k_x(\tau_2,x_2,i_2)\rangle 
= 
\sum_{i_1=1}^2 \left (\delta_{i_1i_2} + |\SS^{\textrm{\tiny (v)}}_{i_1i_2}|^2 \right ) 
\Big[ W(\tau_{12};0,x_2) + W(\tau_{12};x_2,0) \Big] 
\neq 0  \,.
\label{Kirchhoff4}
\ee
Thus, helicity is not conserved across the defect, 
in agreement with the breaking of the $U_{\textrm{\tiny a}}(1)$ symmetry in this phase.

In the axial phase, the two-point functions read
\bea
\label{corr-11-mod-ax}
\langle \chi_1(\tau_1,x_1,i_1)\,\chi_1^*(\tau_2,x_2,i_2)\rangle 
&=&
\langle \chi^\ast_1(\tau_1,x_1,i_1)\,\chi_1(\tau_2,x_2,i_2)\rangle 
=
\delta_{i_ii_2}\, W(\tau_{12};x_1,x_2) 
\\
\label{corr-22-mod-ax}
\langle \chi_2(\tau_1,x_1,i_1)\,\chi_2^*(\tau_2,x_2,i_2)\rangle 
&=&
\langle \chi^\ast_2(\tau_1,x_1,i_1)\, \chi_2(\tau_2,x_2,i_2)\rangle 
=  
\delta_{i_1i_2}\, W(\tau_{12};x_2,x_1) 
\\
\label{corr-mixed1-mod-ax}
\langle \chi_1^*(\tau_1,x_1,i_1)\, \chi_2^*(\tau_2,x_2,i_2)\rangle 
&=&
\overline{\langle \chi_2(\tau_2,x_2,i_2)\, \chi_1 (\tau_1,x_1,i_1)\rangle}
= 
\SS^{\textrm{\tiny (a)}}_{i_1i_2}\, W(\tau_{12};x_1,-x_2)
\\
\label{corr-mixed2-mod-ax}
\langle \chi_1 (\tau_1,x_1,i_1)\, \chi_2 (\tau_2,x_2,i_2)\rangle 
&=&
\overline{ \langle \chi_2^\ast(\tau_2,x_2,i_2)\,\chi_1^*(\tau_1,x_1,i_1)\rangle }
=
\SS^{\textrm{\tiny (a)}*}_{i_1i_2} \;
W(\tau_{12};x_1,-x_2)\,.
\phantom{xxxx}
\eea
These correlators satisfy the KMS conditions 
\bea
\label{KMS1-ax}
\langle \,\chi_{r}(\tau_1,x_1,i_1)\, \chi_{r}^*(\tau_2+\tau+\textrm{i},x_2,i_2)\, \rangle
&=&
\langle \,\chi_{r}^*(\tau_2+\tau, x_2,i_2)\,\chi_{r}(\tau_1,x_1,i_1)\,  \rangle
\\
\label{KMS2-ax}
\langle \,\chi_{r_1}(\tau_1,x_1,i_1)\, \chi_{r_2}(\tau_2+\tau+\textrm{i},x_2,i_2)\, \rangle
&=&
\langle \,\chi_{r_2}(\tau_2+\tau,x_2,i_2)\,\chi_{r_1}(\tau_1,x_1,i_1)\,  \rangle
\\
\label{KMS3-ax}
\langle \,\chi^\ast_{r_1}(\tau_1,x_1,i_1)\, \chi^\ast_{r_2}(\tau_2+\tau+\textrm{i},x_2,i_2)\, \rangle
&=&
\langle \,\chi^\ast_{r_2}(\tau_2+\tau,x_2,i_2)\,\chi^\ast_{r_1}(\tau_1,x_1,i_1)\,  \rangle
\hspace{1cm}
\eea
where $r, r_1, r_2 \in \{1,2\}$.

The electric and helical transport can be studied also in the axial phase, as done above for the vector phase.
In this case the helical current satisfies the Kirchhoff law, 
while the electric current violates this law, 
in agreement with the symmetry content of the axial phase.

\newpage
%%%%%%%%%%%%%%%%%%%%%%%%%%%%%%%%%%%%%%%%%%%
\section{Special bipartitions}
\label{sec_limiting-regimes}

In this section we discuss some limiting regimes of the spatial bipartition shown in Fig.\,\ref{fig-setup}
whose the modular Hamiltonians are local operators.

%%%%%%%%%%%%%%%%%%%%%%%%%%%%%%%%%
\subsection{Two equal intervals at large separation distance}
\label{subsec_limit_distant}

The limiting regime where the equal intervals are at large separation distance can be explored by 
first setting $b=a+\ell$, $x=a+v$ with $v \in [0, \ell]$ and then sending $a\to \infty$.
In this limit we have $\tilde{x} = a + \tilde{v} +O(1/a)$, where $\tilde{v} \equiv \ell - v$.

In this limit, the modular Hamiltonians found in Sec.\,\ref{sec_eh} become local because
the weight functions (\ref{beta-loc-def}) and (\ref{beta-biloc-def}) simplify respectively to
\be
\label{betas-interval-far}
\beta_{\textrm{\tiny loc}}(x) =\beta_0(v) +O(1/a^2)
\;\;\qquad\;\;
\beta_{\textrm{\tiny bi-loc}}(x) =O(1/a)
 \;\;\qquad\;\;
\beta_0(v) \equiv \frac{v(\ell-v)}{\ell}
\ee
where it is worth remarking that $\beta_0(v)$ with $v \in [0, \ell]$ is the weight function occurring in 
the modular Hamiltonian of an interval of length $\ell$ in the infinite line with the first endpoint in the origin,
when the entire system is in its ground state \cite{Hislop:1981uh, Casini:2011kv}.

In this limit, the function (\ref{xi-def}) occurring in the modular flow becomes
\be
 \xi(\tau, x) 
% \,\equiv\, w^{-1}\big( w(x)+ 2\pi\,\tau\big)
\,=\,
a +  \zeta(\tau, v)  + O(1/a)
\;\;\qquad\;\;
\zeta(\tau, v) \equiv \frac{\ell\, v\, e^{2\pi \tau}}{\ell +(e^{2\pi \tau}-1) v}\,.
\ee
By using this result and taking the limit of (\ref{psi-mod-flow-ab}), for the modular flow we obtain
\be
\psi_1(\tau,x,i)  
=
\Big[
\sqrt{ \partial_v \zeta(\tau,v) } \; \psi_1(a+\zeta,i)
\Big]
\! \Big|_{\zeta = \zeta(\tau,v)}
 \quad
  \psi_2(\tau,x,i)  
=
\Big[
\sqrt{ \partial_v \zeta(\tau,v)} \; \psi_2(a+\zeta,i)
\Big]\!
 \Big|_{\zeta = \zeta(-\tau,v)}
\ee
where we used that $ \tfrac{\beta_0(\zeta)}{\beta_0(v)} = \partial_v \zeta(\tau,v)$.

The correlators along the modular flow in the two phases and in this limiting regime
can be written from the expressions in (\ref{corr-11-mod})-(\ref{corr-mixed2-mod}) 
and in (\ref{corr-11-mod-ax})-(\ref{corr-mixed2-mod-ax}),
by observing that (\ref{Wfunc}) in this limit becomes
\be
\label{W-function-interval-far}
W(\tau; x, y) 
=
\frac{\ell}{ 2\pi \textrm{i} \, \big[ v(\ell-y_0)\, e^{\pi \tau} - y_0(\ell-v)\, e^{-\pi \tau} - \textrm{i}  \varepsilon \big]}\,.
\ee

%%%%%%%%%%%%%%%%%%%%%%%%%%%%%%%%%
\subsection{Interval with the defect in its center}
\label{subsec_interval_adj}

The bipartition of the line given by an interval with the defect in its center,
when the massless Dirac field is in its ground state, 
can be studied by taking the limit $a \to 0$ in all the results discussed above,
whenever it is allowed (this is not the case e.g. for the entanglement entropies (\ref{ee-Asym-def})).

In this limit, the function (\ref{w}) becomes
\be
\label{w-func-adjacent}
 w(x) =  \log\left( \frac{x+b}{b-x} \right)
  \;\;\qquad\;\;
 x\in [0,b)
\ee
and the weight functions (\ref{beta-loc-def}) and (\ref{beta-biloc-def}) simplify respectively to
\be
\label{betas-a=0}
\beta_{\textrm{\tiny loc}}(x) \,\to\,\beta_0(x) 
\;\;\qquad\;\;
\beta_{\textrm{\tiny bi-loc}}(x) \,\to\, 0
\;\;\qquad\;\;
\beta_0(x) \equiv \frac{b^2-x^2}{2b} = \frac{1}{w'(x)}\,.
\ee
Thus, in this limit, the modular hamiltonians become a local operator
which has the same form of the modular Hamiltonian of the interval of length $2b$ on the line centered in the origin.
We remark that, 
although the scattering matrix $\SS$ does not appear explicitly in this local operator, 
it enters in the definitions of the Dirac field in the energy density (\ref{T00-lambda-def}).

\begin{figure}[t!]
\vspace{-.8cm}
\hspace{-1.4cm}
%\begin{center}
\includegraphics[width=1.2\textwidth]{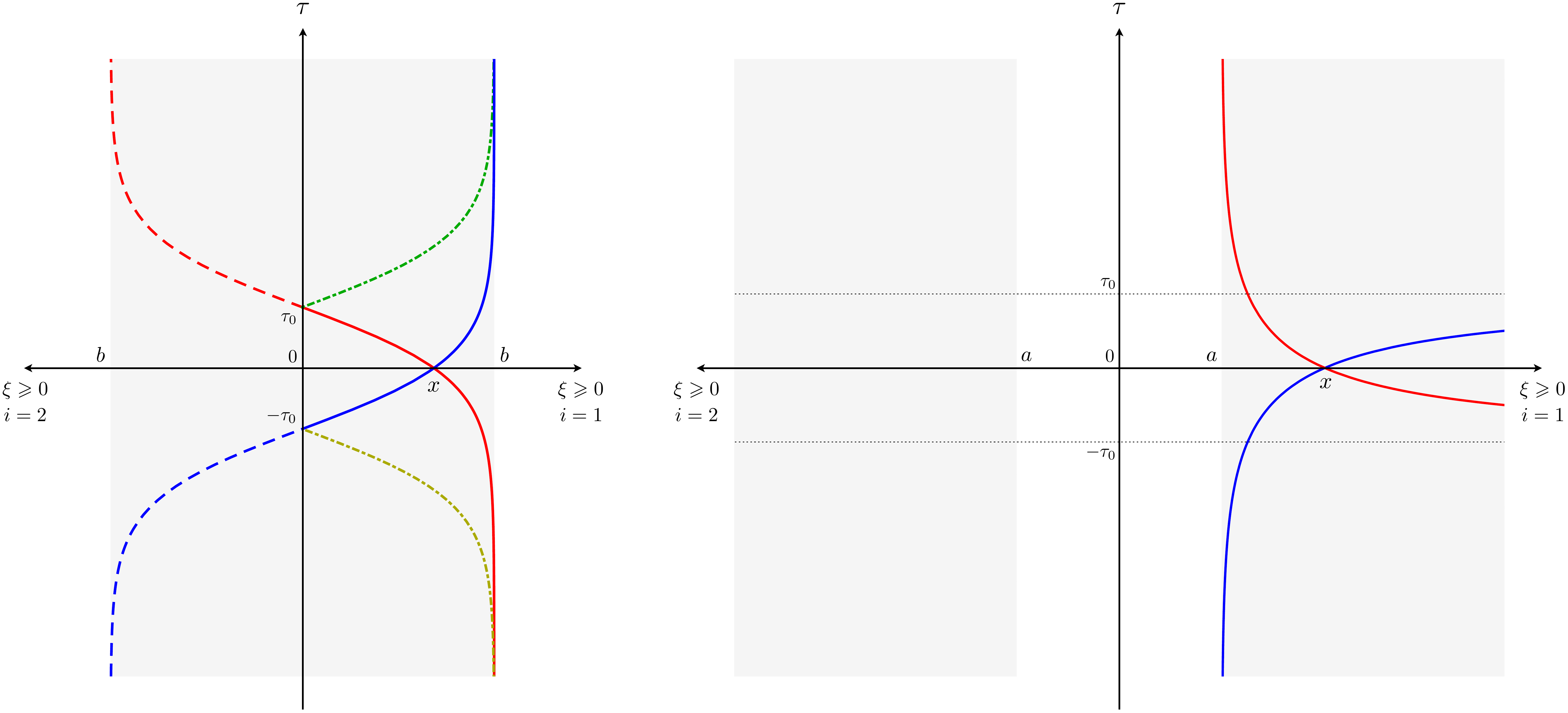}
% \end{center}
\vspace{-.4cm}
\caption{ 
Evolutions of the arguments of the fields along the local modular flows.
Left: Interval of length $2b$ with the defect in its center (see (\ref{xi-mev-adj-1}) and (\ref{xi-mev-adj-2})).
Right: Two semi-infinite lines at the same distance $a$ from the defect (see (\ref{xi-mev-semi-12})).
}
\label{fig-xitau-limits}
\end{figure}

The limit $a\to 0$ of the function $ \xi(\tau, x)$ in (\ref{xi-def}) can be written in terms of $w(x)$  in (\ref{w-func-adjacent}) as 
\be
 \xi(\tau, x) 
% \,\equiv\, w^{-1}\big( w(x)+ 2\pi\,\tau\big)
\,=\,
 b\; \frac{e^{2\pi \tau}  e^{w(x)} -1 }{e^{2\pi \tau}  e^{w(x)} + 1}
 \,=\,
 b\; \frac{x\, \cosh(\pi \tau) + b\, \sinh(\pi \tau)}{b\, \cosh(\pi \tau) + x\, \sinh(\pi \tau)}\,.
\ee

The non-local contributions in the modular flow (\ref{psi-mod-flow-ab}) vanish when $a\to 0$.
However, we remark that $\xi=0$ at some point of its modular evolution where $|\tau| = \tau_0$
(see Fig.\,\ref{fig-xitau-limits}, left panel).
Taking into account the defect boundary condition (\ref{bc3}) at this point of the modular evolution, 
for the modular flow of the components of the Dirac field we find 
\bea
\label{xi-mev-adj-1}
\psi_1(\tau,x,i)  
&=&
\left\{ \begin{array}{ll}
\sqrt{ \partial_x \xi} \;\psi_1(\xi,i)   &  \tau \geqslant -\tau_0
 \\
 \rule{0pt}{.8cm}
 \displaystyle
\sqrt{ \partial_x \xi} \; \sum_{j=1}^2 \SS_{ij}\, \psi_2(\xi,j)  \hspace{.6cm} &  \tau \leqslant  -\tau_0
 \end{array}
 \right.
 \;\;\qquad\;\;
\xi = \xi(\tau,x)
\\
\label{xi-mev-adj-2}
\rule{0pt}{1.3cm}
\psi_2(\tau,x,i)  
&=&
\left\{ \begin{array}{ll}
\sqrt{ \partial_x \xi} \; \psi_2(\xi,i)   &  \tau \leqslant \tau_0
 \\
 \rule{0pt}{.8cm}
 \displaystyle
\sqrt{ \partial_x \xi} \;  \sum_{j=1}^2 \SS^\ast_{ij}\, \psi_1(\xi,j)  \hspace{.6cm} &  \tau \geqslant  \tau_0
 \end{array}
 \right.
 \;\;\qquad\;\;
\xi = \xi(-\tau,x)\,.
\eea
The evolution of the arguments of the fields in the r.h.s.'s are shown in the left panel of Fig.\,\ref{fig-xitau-limits}.
The splitting of the curve at $|\tau| = \tau_0$, where $\xi =0$, is due to the fact that the defect 
allows both reflection and transmission.

When $a \to 0$, the function $W(\tau; x, y) $ is (\ref{Wfunc}) with $w(x)$ given by (\ref{w-func-adjacent}), namely
\be
\label{W-function-a=0}
W(\tau; x, y) 
=
\frac{b}{ \textrm{i}\pi \big[(x+b)(b-y)\, e^{\pi \tau} - (y+b)\, (b-x)\, e^{-\pi \tau} - \textrm{i}  \varepsilon\big]}\,.
\ee
By employing this expression either  in (\ref{corr-11-mod})-(\ref{corr-mixed2-mod}) or in (\ref{corr-11-mod-ax})-(\ref{corr-mixed2-mod-ax}),
we obtain the correlation functions along the modular flow when $a\to 0$, either in the vector phase or in the axial phase respectively.

%\newpage
%%%%%%%%%%%%%%%%%%%%%%%%%%%%%%%%%
\subsection{Two semi-infinite lines}
\label{subsec_limit_semi-infinite}

We find worth considering also the bipartition where
the subsystem is made by two semi-infinite lines at the same distance $a>0$ from the defect,
which can be obtained by taking $b \to +\infty $ in Fig.\,\ref{fig-setup}.
In this limit, the function (\ref{w}) becomes
\be
\label{w-func-semi-infinite}
 w(x) =  \log\! \bigg( \frac{x-a}{x+a} \bigg)
 \;\;\qquad\;\;
 x>a
\ee
and for the weight functions (\ref{beta-loc-def}) and (\ref{beta-biloc-def}) one finds 
\be
\label{betas-b-infty}
\beta_{\textrm{\tiny loc}}(x) =\beta_{0}(x)  +O(1/b)
\qquad
\beta_{\textrm{\tiny bi-loc}}(x) =\frac{\beta_{0}(x)}{x} +O(1/b)
\qquad
\beta_{0}(x) \equiv
\frac{x^2-a^2}{2a} \,.
% \qquad
%b\to \infty
\ee
We remark that, 
despite the fact that the limit of $\beta_{\textrm{\tiny bi-loc}}(x)$ is non vanishing,
the modular Hamiltonian becomes local in the same limit
because the fields in the bi-local term (\ref{K_A-bilocal-def}) vanish, as also discussed in \cite{Mintchev:2020uom}.
Indeed, $\tilde{x} \to \infty$ when $b \to +\infty $, hence $\psi_j(\tilde{x})\to 0$ for $j\in \{1,2\}$.

In this limiting regime,  the function (\ref{xi-def}) simplifies to
\be
 \xi(\tau, x) 
\,=\,
 -\, a\; \frac{e^{w(x)+2\pi \tau} + 1 }{e^{w(x)+2\pi \tau} - 1}
 \,=\,
 a\, \frac{x\, \cosh(\pi \tau) -a\, \sinh(\pi \tau)}{a\, \cosh(\pi \tau) -x \, \sinh(\pi \tau)}\,.
\ee
This expression provides the modular flow of the Dirac field, 
which can be found by taking the limit $b \to +\infty $ of (\ref{psi-mod-flow-ab}).
The result reads
\be
\label{xi-mev-semi-12}
\psi_1(\tau,x,i)  
=
\Big[ 
\sqrt{ \partial_x \xi} 
\; \psi_1(\xi,i)   \Big]\! \Big|_{\xi = \xi(\tau,x)}
\;\;\qquad\;\;
\psi_2(\tau,x,i)  
=
\Big[ 
\sqrt{ \partial_x \xi} 
\; \psi_2(\xi,i)   \Big]\! \Big|_{\xi = \xi(-\tau,x)}
\ee
where the first expression holds for $\tau \leqslant \tau_0$ and the second one for $\tau \geqslant -\tau_0$,
with $\tau_0 = \tfrac{1}{2\pi} |w(x)|$.
In the right panel of Fig.\,\ref{fig-xitau-limits}, we show the evolution of the arguments of the fields
in the r.h.s.'s of (\ref{xi-mev-semi-12}) for a given point which has spatial coordinate $(x,1)$ at $\tau=0$.

The correlators along the modular flow in this limiting regime
can be written by first taking the limit $b \to +\infty $ of (\ref{Wfunc}), that gives
\be
\label{W-function-b=infty}
W(\tau; x, y) 
=
\frac{a}{\pi \textrm{i}\, \big[ (x-a)(y+a)\, e^{\pi \tau} - (y-a)\, (x+a)\, e^{-\pi \tau} - \textrm{i}  \varepsilon \big]}
\ee
and then employing the resulting expression either in 
(\ref{corr-11-mod})-(\ref{corr-mixed2-mod}) for the vector phase 
or in (\ref{corr-11-mod-ax})-(\ref{corr-mixed2-mod-ax}) for the axial phase. 

Considering the partition $A\cup B$ of the line where $A$ is the interval with the defect in its center and $B$ its complement,
the local modular Hamiltonians obtained in Sec.\,\ref{subsec_interval_adj} and in this subsection can be combined 
to construct the full modular Hamiltonian 
\be
K_{A\cup B} \,=\, K_A \otimes \boldsymbol{1}_B - \boldsymbol{1}_A \otimes K_B
\ee
where $\boldsymbol{1}_A$ and $\boldsymbol{1}_B$ denote the identity operators on $A$ and $B$ respectively.

%%%%%%%%%%%%%%%%%%%%%%%%%%%%%%%%%%%%%%%%%%%%%%
%\newpage

\section{Modular evolution in the spacetime}
\label{space-time}

The modular evolution and the modular correlation functions of the fields at fixed time $t=0$
has been considered in Sec.\,\ref{sec_mod_flow} and Sec.\,\ref{sec_correlators}.
In the following analysis, we extend these results to generic values of the physical time $t$
by taking advantage of the fact that, even in the presence of the defect, 
in both the phases the Dirac field depends on the light cone coordinates defined by 
\be
(u_\pm ,i) = (x\pm t, i)  \;\;\qquad\;\; 
x\geqslant 0 
\;\;\qquad\;\; i\in \{1,2\}\, .  
\label{lc1}
\ee

The canonical anti-commutation relations in the algebras $\cal A_+$ and $\cal B_+$ imply 
%the following light cone anti-commutators
\bea
\label{A1}
& & \hspace{-.5cm}
[\psi_1 (u_+,i)\, ,\, \psi^*_1 (v_+,j)]_+ = \delta_{ij} \,\delta (u_+ - v_+) 
\hspace{.73cm}
[\psi_2 (u_-,i) \,, \,\psi^*_2 (v_-,j)]_+ = \delta_{ij} \,\delta (u_- - v_-) 
\phantom{xxxxx}
\\
\label{A2}
& & \hspace{-.5cm}
[\psi_1 (u_+,i)\, ,\, \psi^*_2 (v_-,j)]_+=\SS_{ij}\, \delta (u_+ + v_-)
\hspace{.7cm}
[\psi_2 (u_-,i)\, ,\, \psi^*_1 (v_+,j)]_+=\SS^*_{ij}\, \delta (u_- + v_+)\, .  
\eea

In light cone coordinates, the auxiliary fields are introduced as follows
\begin{equation}
\widetilde{\psi}_1 (u_+, i) = \sum_{j=1}^2 \U_{ij}\, \psi_1 (u_+,j)
\;\;\qquad\;\;
\widetilde{\psi}_2 (u_-, i) = \sum_{j=1}^2 \U_{ij}\, \psi_2 (u_-,j)\,.
\label{auxlc}
\end{equation}

\begin{figure}[t!]
\vspace{-.8cm}
\hspace{-1.2cm}
%\begin{center}
\includegraphics[width=1.17\textwidth]{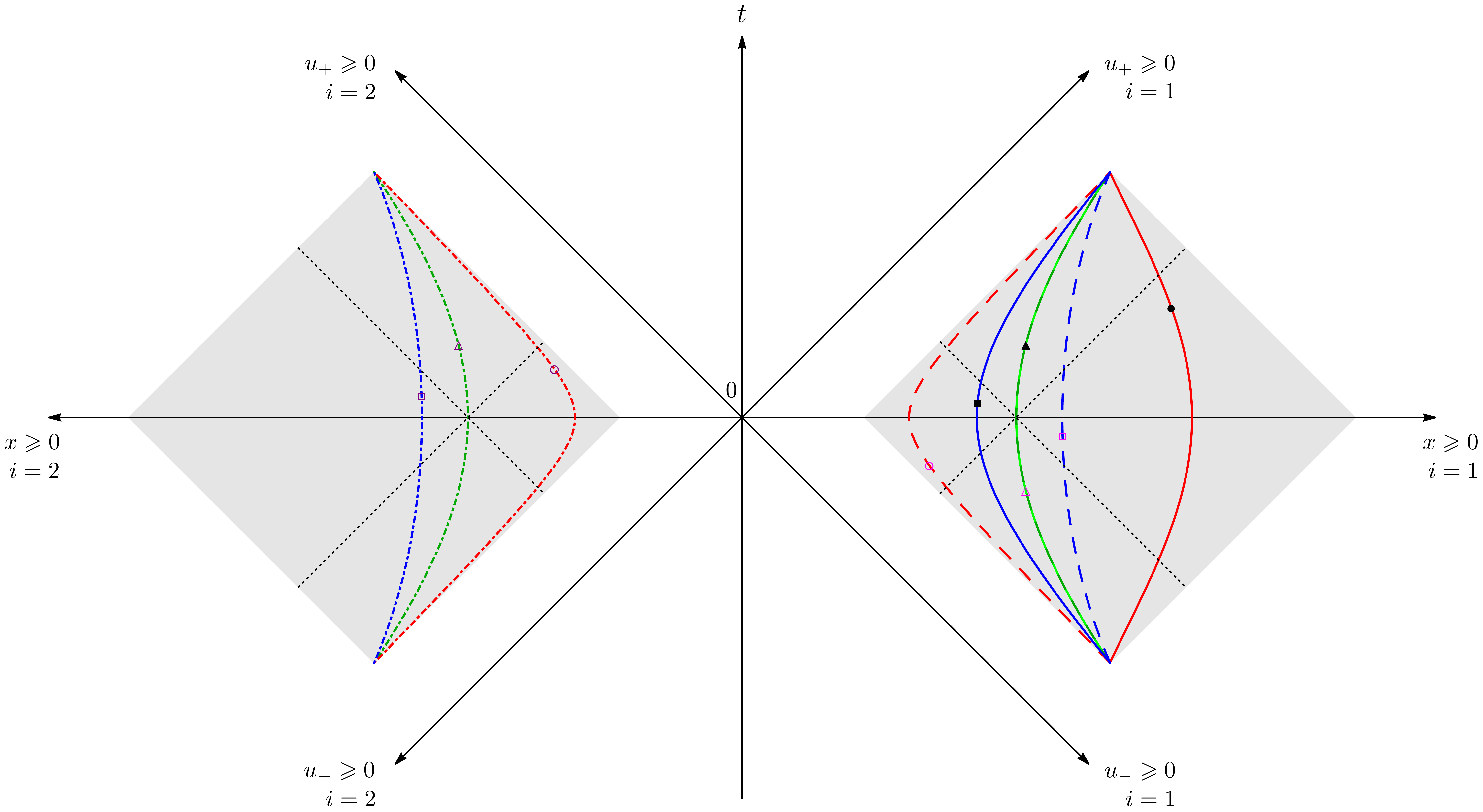}
% \end{center}
\vspace{-.0cm}
\caption{ 
Three sets of conjugated modular trajectories in the spacetime
for the modular flow of the Dirac field in the presence of a defect in the origin (see (\ref{psi-mod-flow-phys-lc})).
%Each sets corresponds to a colour and the three modular trajectories within each sets are indicated through different lines.
}
\label{figure-diamond}
\end{figure}

Let us consider the spacetime region defined by (see the grey region in Fig.\,\ref{figure-diamond})
\be 
\mathcal{D} \equiv \mathcal{D}_{1} \cup \mathcal{D}_{2}
\;\;\qquad\;\;
\mathcal{D} _{i}=
\big\{ \big( (u_+,i),(u_-,i)\big) : a\leqslant u_\pm \leqslant b\, \big\} \,.
\label{domain}
\ee

By  applying the results of \cite{Mintchev:2020uom} for the modular Hamiltonians in the spacetime
to the auxiliary fields $\{\widetilde{\psi}_1(u_+,i), \widetilde{\psi}_2(u_-,i)\}$ for $i=1,2$, 
we obtain
\be
\label{K_A-decomposition-lc}
K  \,= \, K^{\textrm{\tiny loc}} + K^{\textrm{\tiny bi-loc}} \,.
\ee 
The local term $K^{\textrm{\tiny loc}} $ in this decomposition reads
\be
\label{K_A-local-def-lc}
K^{\textrm{\tiny loc}} 
\,=\,
2\pi  \sum_{i=1}^2 \int_a^b \! \beta_{\textrm{\tiny loc}}(u) \, \widetilde{T}_{\textrm{\tiny loc}}(0,u,i) ,\rd u
\,=\,
2\pi  \sum_{i=1}^2 \int_a^b \! \beta_{\textrm{\tiny loc}}(u) \, T_{tt}(0,u,i)\,\rd u
\ee
where (\ref{T00-lambda-def-tilde}) and (\ref{T00-lambda-def}) have been used.
The bi-local term in (\ref{K_A-decomposition-lc}) is
\be
\label{K_A-bilocal-def-lc}
K^{\textrm{\tiny bi-loc}}  
\,=\,
2\pi 
\sum_{i=1}^2 \int_a^b \! \beta_{\textrm{\tiny bi-loc}}(u) 
\, \widetilde{T}_{\textrm{\tiny bi-loc}}(0,u, \tilde{u},i)  \, \rd u
\,=\,
2\pi 
\sum_{i=1}^2 \int_a^b \! \beta_{\textrm{\tiny bi-loc}}(u) 
\, T_{\textrm{\tiny bi-loc}}(0,u, \tilde{u},i)  \, \rd u
\ee
where the bi-local operators (\ref{T-bilocal-def}) and (\ref{T-bilocal-defphys}) have been employed
and $\tilde{u}_\pm \equiv a b/ u_\pm$ is conjugate to $u_\pm$.
The weight functions $\beta_{\textrm{\tiny loc}}(u)$ and $\beta_{\textrm{\tiny bi-loc}}(u)$
are (\ref{beta-loc-def}) and (\ref{beta-biloc-def}) respectively.

The modular flow of the auxiliary fields  is governed by the following initial value problems
\bea
\label{lc-mod-evolution1}
\textrm{i}\,\frac{ d\widetilde{\psi}_1(\tau,u_+,i)}{d\tau}  = \big[\,K \,, \widetilde{\psi}_1(\tau,u_+,i)\,\big]_{-}
\;\; \qquad \;\; \widetilde{\psi}_1(0,u_+,i) = \widetilde{\psi}_1(u_+,i) 
\\
\label{lc-mod-evolution2}
\textrm{i}\,\frac{ d\widetilde{\psi}_2(\tau,u_-,i)}{d\tau}  = \big[\,K \,, \widetilde{\psi}_2(\tau,u_-,i)\,\big]_{-}
\;\; \qquad \;\;
\widetilde{\psi}_2(0,u_-,i) = \widetilde{\psi}_2(u_-,i) 
\eea
where  the initial configurations $\widetilde{\psi}_1(u_+,i)$ and $\widetilde{\psi}_2(u_-,i)$ 
are related to the initial configurations of the physical fields through (\ref{auxlc}). 
The system made by the four partial differential equations in (\ref{lc-mod-evolution1}) and (\ref{lc-mod-evolution2}) 
decouples into two independent systems corresponding to $i=1$ and $i=2$,
each of them made by two partial differential equations.
These equations are of the form analysed and solved in \cite{Mintchev:2020uom}. 
By employing the solution found in \cite{Mintchev:2020uom}, in this case
for the modular flow of the auxiliary fields we find 
\begin{equation}
\label{lc4}
\left\{ \begin{array}{l}
\displaystyle
\widetilde{\psi}_1(\tau,u_+,i)
=
\Bigg[
P(\xi;  u_+) 
\Bigg(
 \big( a \,b + \xi \,u_+ \big) \,\widetilde{\psi}_1(\xi,i) 
- \frac{a\,b}{\xi}\; e^{\textrm{i}\alpha_i} \big(\xi - u_+\big)\, \widetilde{\psi}_2(ab/\xi,i)
\Bigg)
\Bigg]
\! \Bigg|_{\xi = \xi(\tau,u_+)}
\\
\rule{0pt}{.85cm}
\displaystyle
\widetilde{\psi}_2(\tau,u_-,i)
=
\Bigg[
P(\xi;  u_-) 
\Bigg(
 \big( a \,b + \xi \,u_- \big) \,\widetilde{\psi}_2(\xi,i) 
- \frac{a\,b}{\xi}\; e^{-\textrm{i}\alpha_i} \big(\xi - u_-\big)\, \widetilde{\psi}_1(ab/\xi,i)
\Bigg)
\Bigg]
\! \Bigg|_{\xi = \xi(-\tau,u_-)}
\end{array}
\right. .
\end{equation}
This solution extends (\ref{psi-mod-flow-ab0}) to generic $t$ in terms of the light coordinates $u_\pm$.

The modular flow of the physical fields $\psi_r$ can be found by
inverting (\ref{auxlc}) and employing the modular flow (\ref{lc4}) for the auxiliary fields.
The result is
\be
\label{psi-mod-flow-phys-lc}
\hspace{-.1cm}
\left\{ \begin{array}{l}
\displaystyle
\psi_1(\tau,u_+,i) = 
\Bigg[
P(\xi;  u_+) 
\Bigg(
 \big( a \,b + \xi u_+ \big ) \,\psi_1(\xi,i) 
- \frac{a\,b}{\xi} \big(\xi - u_+\big)\, \sum_{j=1}^2 \SS_{ij}\, \psi_2(ab/\xi,j)
\Bigg)
\Bigg]
\! \Bigg|_{\xi = \xi(\tau,u_+)}
\\
\rule{0pt}{.9cm}
\displaystyle
\psi_2(\tau, u_-,i)
=
\Bigg[
P(\xi;  u_-) 
\Bigg(
 \big( a \,b + \xi  u_- \big ) \,\psi_2(\xi,i) 
- \frac{a\,b}{\xi} \big(\xi - u_-\big)\, \sum_{j=1}^2 \SS^{*}_{ij}\, \psi_1(ab/\xi,j)
\Bigg)
\Bigg]
\! \Bigg|_{\xi = \xi(-\tau,u_-)}
\end{array}
\right. .
\ee

This flow has the features highlighted for (\ref{psi-mod-flow-ab}).
In particular, the modular evolution of each component in (\ref{psi-mod-flow-phys-lc}) 
is obtained by combining
the modular evolutions of the initial data for three fields whose arguments follow
different trajectories  in general. 
The initial points at $\tau=0$ of
these trajectories are related by conjugation (see (\ref{x-conjugate})).
Furthermore, the scattering matrix $\SS$ characterising the defect explicitly occurs in the mixing described by (\ref{psi-mod-flow-phys-lc}).

In Fig.\,\ref{figure-diamond} we show three sets of conjugated modular trajectories in the spacetime,
denoting them through different colours. 
The three modular trajectories within each set are indicated by different kind of lines (solid, dashed and dashed-dotted)
and their initial points at $\tau=0$ correspond to the markers characterised by the same kind of symbol 
(circle, square or triangle).
The filled markers indicate the initial points with coordinates $P_0= ((u_{+,0}, i), (u_{-,0}, i))$
(in particular, $i=1$ in Fig.\,\ref{figure-diamond}), 
while the empty markers correspond to the points obtained from $P_0$ through the conjugation (\ref{x-conjugate}), 
namely $((ab/u_{+,0}, i), (ab/u_{-,0}, i))$ and $((ab/u_{+,0}, j), (ab/u_{-,0}, j))$, with $j\neq i$.
The green set of modular trajectories is characterised by the fact that its curves pass through the points 
$((\sqrt{ab}\,, i), (\sqrt{ab}\,, i))$ with $i \in \{1,2\}$. 
In this case, the two modular trajectories within the same grey diamond coincide.

In the vector phase, the modular flow of the Dirac field can be found by employing (\ref{psiu}) into (\ref{psi-mod-flow-phys-lc}).
The result reads
\be
\label{psi-mod-flow-vector-lc}
\hspace{-.3cm}
\left\{ \begin{array}{l}
\displaystyle
\lambda_1(\tau,u_+,i) = 
\left[
P(\xi;  u_+) 
\left(
 \big( a \,b + \xi u_+ \big ) \,\lambda_1(\xi,i) 
- \frac{a\,b}{\xi} \big(\xi - u_+\big)\, \sum_{j=1}^2 \SS^{\textrm{\tiny (v)}}_{ij}\, \lambda_2(ab/\xi,j)
\right)
\right]
\! \bigg|_{\xi = \xi(\tau,u_+)}
\\
\rule{0pt}{1.1cm}
\displaystyle
\lambda_2(\tau, u_-,i)
=
\left[
P(\xi;  u_-) 
\left(
 \big( a \,b + \xi  u_- \big ) \,\lambda_2(\xi,i) 
- \frac{a\,b}{\xi} \big(\xi - u_-\big)\, \sum_{j=1}^2 \SS^{\textrm{\tiny (v)}*}_{ij}\, \lambda_1(ab/\xi,j)
\right)\right]
\! \bigg|_{\xi = \xi(-\tau,u_-)}
\end{array}
\right. .
\ee
These fields provide the correlation functions along the modular flow in the vector phase, in the light cone coordinates. 
They read
\bea
\label{corr-11-mod-lc}
\langle \lambda_1(\tau_1,u_{1+},i_1)\,\lambda_1^*(\tau_2,u_{2+},i_2)\rangle 
&=&
\langle \lambda^\ast_1(\tau_1,u_{1+},i_1)\,\lambda_1(\tau_2,u_{2+},i_2)\rangle 
=
\delta_{i_1i_2}W(\tau_{12};u_{1+},u_{2+}) \qquad  \quad 
\\
\label{corr-22-mod-lc}
\langle \lambda_2(\tau_1,u_{1-},i_1)\,\lambda_2^*(\tau_2,u_{2-},i_2)\rangle 
&=&
\langle \lambda^\ast_2(\tau_1,u_{1-},i_1)\,\lambda_2(\tau_2,u_{2-},i_2)\rangle 
=  
\delta_{i_1i_2} W(\tau_{12};u_{2-},u_{1-}) \qquad \quad 
\\
\label{corr-mixed1-mod-lc}
\langle \lambda_1(\tau_1,u_{1+},i_1)\,\lambda_2^*(\tau_2,u_{2-},i_2)\rangle 
&=&
\overline{\langle \lambda_2(\tau_2,u_{2-},i_2)\,\lambda_1^*(\tau_1,u_{1+},i_1)\rangle}
= 
\SS^{\textrm{\tiny (v)}}_{i_1i_2}W(\tau_{12};u_{1+},-u_{2-})  \qquad \qquad 
\\
\label{corr-mixed2-mod-lc}
\langle \lambda_1^\ast(\tau_1,u_{1+},i_1)\,\lambda_2(\tau_2,u_{2-},i_2)\rangle 
&=&
\overline{ \langle \lambda_2^\ast(\tau_2,u_{2-},i_2)\,\lambda_1(\tau_1,u_{1+},i_1)\rangle }
=
\SS^{\textrm{\tiny (v)}*}_{i_1i_2} W(\tau_{12};u_{1+},-u_{2-})\,.
\eea 

In the axial phase, expressions analogous to (\ref{psi-mod-flow-vector-lc})-(\ref{corr-mixed2-mod-lc}) can be written.

%\newpage
%%%%%%%%%%%%%%%%%%%%%%%%%%%%%%%%%%%%%%%%%%%
\section{Conclusions}
\label{sec_conclusions}

In this manuscript we have studied some modular Hamiltonians and the corresponding modular flows
for the massless Dirac field on a line in the presence of a defect characterised by a $2 \times 2$ unitary scattering matrix $\SS$.
The system is in its ground state and
the bipartition of the line is given by the union of two disjoint equal intervals at the same distance from the defect. 
For preventing energy dissipation, the defect which allows for both reflection and transmission, must satisfy the Kirchhoff rule (\ref{K}).
This leads to two inequivalent models, the vector phase and the axial phase,
characterised by the scale invariant boundary conditions (\ref{bc1}) and (\ref{bc2}) respectively,
where different symmetries are preserved, as discussed in Sec.\,\ref{sec_generals}.

By employing a basis of auxiliary fields (see Sec.\,\ref{sec_aux_fields}) and the results of \cite{Mintchev:2020uom},
we have obtained the modular Hamiltonians (\ref{K_A-decomposition}),
where the local term is given by (\ref{K_A-local-def})-(\ref{T00-lambda-def})
and the bi-local term by (\ref{K_A-bilocal-def})-(\ref{T-bilocal-defphys}).
The bi-local operator (\ref{T-bilocal-defphys}) depends explicitly on the scattering matrix characterising the defect.
Furthermore, considering the integrands in the sum of bi-local terms (\ref{K_A-bilocal-def}), 
for any point, two other conjugate points are also involved.
This feature, which is due to the fact that both reflection and transmission are allowed by the defect,
represents an important difference with respect to the non-local modular Hamiltonians for the massless Dirac field 
available in the literature, where either two or infinitely many conjugate points are involved 
 \cite{Casini:2009vk, Klich:2015ina, Blanco:2019xwi, Fries:2019ozf,Mintchev:2020uom}.
The symmetry of the bipartition and the nature of the defect 
lead to entanglement entropies that are independent of the scattering matrix (see (\ref{ee-Asym-def})).

The modular flows of the Dirac field generated by these modular Hamiltonians have beed found. 
They are given by the solution (\ref{psi-mod-flow-ab}),
which becomes (\ref{psi-mod-flow-vector}) in the vector phase and (\ref{psi-mod-flow-axial}) in the axial phase. 
These modular flows mix three modular trajectories, as shown also in Fig.\,\ref{fig-xi} and Fig.\,\ref{figure-diamond}.
The correlators of the Dirac field along the modular flows have been written in terms of the function (\ref{Wfunc}), 
where $w(x)$ is given by (\ref{w}). 
Their explicit expressions are (\ref{corr-11-mod})-(\ref{corr-mixed2-mod}) in the vector phase and (\ref{corr-11-mod-ax})-(\ref{corr-mixed2-mod-ax}) in the axial phase.
The modular flow equations lead to write modular equations of motions for these correlators (see e.g. (\ref{modeqdef})).
We have checked that the current-current correlators satisfy the Kirchhoff law at the defect (see e.g. (\ref{Kirchhoff3})).
In some limiting cases for the spatial bipartition, the modular Hamiltonians become local. 
These limits, which have been explored in Sec.\,\ref{sec_limiting-regimes},
correspond to equal intervals at large separation distance, 
to a single interval with the defect in its center or to two semi-infinite lines at the same distance from the defect.

The modular Hamiltonians and the corresponding modular flows found for $t=0$ have been extended in the spacetime to a generic value of 
the physical time $t$ in Sec.\,\ref{space-time}.
The results are given by (\ref{K_A-decomposition-lc})-(\ref{K_A-bilocal-def-lc}) for the modular Hamiltonians and by (\ref{psi-mod-flow-phys-lc}) for the modular flows.
In the vector phase, the explicit expressions for the modular flow of the Dirac field and its correlators are given in (\ref{psi-mod-flow-vector-lc}) and (\ref{corr-11-mod-lc})-(\ref{corr-mixed2-mod-lc})
respectively.

Various directions can be explored in the future. 
For instance, it is natural to consider spatial bipartitions made by an arbitrary number of disjoint intervals in a generic configuration
with respect to the defect, as done in some models on the line without the defect 
\cite{Casini:2009sr, Casini:2009vk, Calabrese:2009ez, Calabrese:2010he,Coser:2013qda, DeNobili:2015dla}.
Some entanglement Hamiltonians in free models on the lattice  \cite{Peschel:2003rdm, Casini:2009sr, EislerPeschel:2009review, Eisler:2017cqi, Eisler:2018ugn}
and also their continuum limits have been studied  \cite{Arias:2016nip,Eisler:2019rnr, DiGiulio:2019cxv}.
It would be interesting to recover also the modular Hamiltonians found here through these lattice calculations. 

Entanglement quantifiers closely related to the modular Hamiltonians are the
corresponding entanglement spectra \cite{Peschel:2003rdm,Casini:2009sr, EislerPeschel:2009review, Li:2008kda}
and entanglement contours for the entanglement entropies \cite{ChenVidal2014,Coser:2017dtb}.
In $1+1$ conformal field theories, some modular Hamiltonians and their entanglement spectra 
have been explored through boundary conformal field theory methods 
\cite{Lauchli:2013jga, Cardy:2016fqc, Tonni:2017jom, Alba:2017bgn,DiGiulio:2019lpb,DiGiulio:2019cxv, Surace:2019mft,Roy:2020frd}
and it is worth trying to employ these techniques  also in the presence of defects.

It would be interesting to study modular Hamiltonians for models where the scale invariance is broken,
by the impurity (see e.g. \cite{AffleckLaFlorencie07, AffleckLaFlorencie09, 
EislerPeschel:2010def, Calabrese:2011ru, EislerPeschel:2012def, Saleur:2013pva, Ossipov14,Bellazzini:2008fu, Casini:2016fgb, Kobayashi:2018lil}, 
where mainly the entanglement entropies have been studied)
or in the bulk e.g. through a mass term \cite{Arias:2016nip, Eisler:2020lyn}.
Modular Hamiltonians in the presence of defects in higher dimensional models \cite{Jensen:2013lxa, Casini:2018nym}
and in the context of the gauge/gravity correspondence \cite{Bachas:2001vj, Azeyanagi:2007qj, Erdmenger:2013dpa, Headrick:2014cta, Jafferis:2015del, Takayanagi:2011zk, Fujita:2011fp, Nozaki:2012qd,  Seminara:2017hhh, Seminara:2018pmr}
deserve further studies.

%\newpage
%%%%%%%%%%%%%%%%%%%%%%%%%%%%
\vskip 30pt 
\centerline{\bf Acknowledgments} 
\vskip 10pt

We are grateful to Paolo Acquistapace, Ra\'ul Arias, Stefano Bianchini, Viktor Eisler and Giuseppe Mussardo 
for insightful discussions and correspondence.

%\newpage

%%%%%%%%%%%%%%%%%%%%%%%%%%%%%%%%%%%%%%%%%%%%%%%%%%%%%
%\newpage

\bibliographystyle{nb}

\bibliography{refsdef}

\end{document}